\documentclass[a4paper,11pt]{article}
\pdfoutput=1 
\usepackage{jheppub} 
\usepackage[T1]{fontenc} 
\usepackage{epsfig,amsmath}
\usepackage{hepnames,hepunits}
\usepackage{booktabs}
\usepackage{bm}
\usepackage{comment}

\usepackage{accents}
\newlength{\dhatheight}
\newcommand{\doublehat}[1]{%
    \settoheight{\dhatheight}{\ensuremath{\hat{#1}}}%
    \addtolength{\dhatheight}{-0.35ex}%
    \hat{\vphantom{\rule{1pt}{\dhatheight}}%
    \smash{\hat{#1}}}}

\usepackage{orcidlink}

\title{Bin-to-bin correlations in the extraction of 
proton's transverse  structure}

\author[a]{L.~Favart\,\orcidlink{0000-0003-1645-7454},}
\emailAdd{laurent.favart@ulb.be}
\author[b]{F.~Hautmann,}
\emailAdd{francesco.hautmann@physics.ox.ac.uk}
\author[c]{A.~Lelek\,\orcidlink{0000-0001-5862-2775},}
\emailAdd{aleksandra.lelek@uantwerpen.be}
\author[d]{L.~Moureaux\,\orcidlink{0000-0002-2310-9266}}
\emailAdd{louis.moureaux@cern.ch}

\affiliation[a]{IIHE, Universit{\' e} Libre de Bruxelles, B 1050 Brussels, Belgium}
\affiliation[b]{University of Oxford, Oxford OX1 3PU, United Kingdom}
\affiliation[c]{Elementaire Deeltjes Fysica, Universiteit Antwerpen, B 2020 Antwerpen, Belgium}
\affiliation[d]{Institut f{\" u}r Experimentalphysik, Universit{\" a}t Hamburg, DE 22761 Hamburg, Germany}

\abstract{We investigate the determination of non-perturbative
QCD contributions, associated with proton's  transverse structure, 
from high-precision experimental data for electroweak  boson
production at the LHC. We demonstrate that these determinations 
are strongly affected by correlated uncertainties across 
transverse-momentum bins in the region of low boson's 
transverse momenta. We illustrate our results through 
a detailed analysis of LHCb measurements of $Z$-boson spectra 
at forward rapidities, and a comparison with 
CMS measurements at central rapidities. We further 
examine the influence on the extraction of the non-perturbative 
parameters from final-state electromagnetic radiation, 
photon-induced background, large-$x$ contributions to 
parton distributions in the forward region, matching 
to   perturbative 
next-to-leading order. We comment on the 
comparison of our results with those from the 
underlying-event tuning of 
parton-shower Monte Carlo event generators.} 


\begin{document} \maketitle
\flushbottom
\unboldmath

\section{Introduction}\label{sec1}

The transverse structure of  hadrons in 
high-energy collisions, 
complementing the longitudinal (collinear) 
 structure and 
providing a full three-dimensional (3D) 
picture~\cite{Angeles-Martinez:2015sea,Abdulov:2021ivr} 
in terms of partonic degrees of freedom,  
 plays an important role for many aspects 
of the physics program at 
present colliders (LHC, RHIC) as well as future 
colliders (EIC~\cite{Proceedings:2020eah}, 
EIC-China~\cite{Xiao:2026tbs},  
HL-LHC~\cite{Azzi:2019yne,LHeC:2020van}, 
FCC~\cite{FCC:2018byv}, 
CEPC~\cite{CEPCPhysicsStudyGroup:2022uwl}). 

On one hand, 
transverse hadron structure gives 
significant non-perturbative QCD 
contributions to the production spectra of many 
final states in the region of 
low transverse momenta, including, 
e.g., electroweak 
bosons~\cite{Angeles-Martinez:2015sea},  
 Higgs bosons~\cite{Cipriano:2013ooa}, 
 heavy flavors~\cite{Boer:2024ylx}.  
 These  contributions affect, for instance,  
 precision determinations of 
Standard Model parameters, such as  
the $W$ boson mass~\cite{CMS:2024lrd}  
and the strong coupling 
$\alpha_s$~\cite{ATLAS:2023lhg,Camarda:2022qdg,Billis:2024dqq}.

  On the other hand,  transverse hadron structure 
 also influences physics  away from 
  the low transverse-momentum region,   
  through   perturbative QCD evolution and   
  transverse-momentum broadening phenomena. 
This is the case, for example, for the physics 
of final states with large multiplicities of 
hard jets~\cite{BermudezMartinez:2021lxz,BermudezMartinez:2021zlg,BermudezMartinez:2022bpj} (due to 
renormalization-group evolution to high mass scales) 
and for the physics of high gluon densities 
at small $x$ and in large nuclei~\cite{Caucal:2025mth,Caucal:2025xxh,Caucal:2024bae} 
(due to high-energy evolution).   
 
In the last couple of years, by exploiting the 
great precision of  the measurements of the 
transverse momentum  of electroweak 
gauge bosons at the Large Hadron 
Collider (LHC) through 
Drell-Yan (DY) lepton-pair decay channels,  
and their sensitivity to 3D hadron structure, 
progress has 
been achieved in the determination of 
non-perturbative 
transverse momentum dependent (TMD) 
parton distribution functions from fits to 
experimental DY data. Extractions of 
TMD sets have been carried out in 
Refs.~\cite{Bacchetta:2022awv,Bacchetta:2024qre,Bury:2022czx,Moos:2025sal,Barry:2023qqh,Barry:2025glq} 
using the framework of analytic TMD 
evolution~\cite{Collins:1984kg,Aybat:2011zv} (CSS), 
and in Refs.~\cite{Bubanja:2023nrd,Zhan:2024lym} 
using  
the framework of 
parton branching (PB) TMD 
evolution~\cite{Hautmann:2017xtx,Hautmann:2017fcj}. 
An extraction of 
the ``intrinsic transverse momentum'' 
$k_T$, controlling the TMD distribution for 
low mass scales in the QCD evolution, 
has been carried out in Ref.~\cite{CMS:2024goo} 
from the tuning of parton-shower 
Monte Carlo event generators~\cite{Sjostrand:2014zea,Bellm:2015jjp} to experimental DY data.  

The comparison of different TMD extractions 
has been the subject of many 
recent discussions in the literature. It is aided by 
computational tools such as the web-based 
library~\cite{Abdulov:2021ivr,Hautmann:2014kza}, which is 
constructed in the spirit of using 
parton-model-like formulas for 
physical cross sections (similarly to the collinear 
case) even in the 
presence of transverse-structure effects.  
Current  issues in TMD determinations 
and non-perturbative effects are examined  
in Refs.~\cite{Cerutti:2026apy,Zaccheddu:2026pyy,Fernando:2025xzv,Camarda:2025lbt,Aslan:2024nqg}.
The  difference in the  
energy behavior of the intrinsic-$k_T$ distributions  
obtained in the parton-shower 
Monte Carlo study~\cite{CMS:2024goo} 
and in the PB TMD study~\cite{Bubanja:2023nrd} 
is investigated specifically 
in Refs.~\cite{Moureaux:2025cyi,Hautmann:2025fkw,Bubanja:2024puv}. 
 Besides TMD distributions, the discussion of 
  non-perturbative effects also 
  involves the non-perturbative components 
 of Sudakov form factors~\cite{Hautmann:2020cyp,Hautmann:2021ovt}, embodied by 
the Collins-Soper~\cite{Collins:1981uk,Collins:1981va} rapidity-evolution kernel. For this kernel, important 
inputs are provided by lattice QCD 
calculations~\cite{Francis:2026czb,Avkhadiev:2026xyf,Tan:2025ofx,Bollweg:2025iol,Bollweg:2025ecn,Avkhadiev:2024mgd,Avkhadiev:2025wps,LPC:2022ibr,Deng:2022gzi,LatticePartonLPC:2023pdv} as well. 

Given the relevance of understanding 
potential differences in current extractions of 
3D proton structure 
from precision DY measurements, 
in this work we carry out a detailed 
study of the influence 
of correlated uncertainties 
across transverse momentum bins  
on such extractions. We illustrate that 
taking into account transverse-momentum 
correlations is essential in order to fully 
exploit the power of precision data in 
TMD determinations. Conversely, we 
point out that overlooking such correlations 
can lead to biases in the shape of the 
TMD distributions and distortions 
in the extracted intrinsic-$k_T$. 

For this study, we focus on the 
LHCb measurements~\cite{LHCb:2021huf} 
of $Z$-boson production at 13\,TeV 
in the forward rapidity region, 
compared  with the analogous 
CMS measurements~\cite{CMS:2022ubq}  
in the central rapidity region. 
While the 
CMS paper~\cite{CMS:2022ubq} 
reports full covariance matrices, the 
LHCb paper~\cite{LHCb:2021huf} does not; we 
then use all the information available from the 
corresponding HepData record on 
statistical and efficiency uncertainties, 
and corresponding correlations,  
 to reconstruct covariance matrices.   
 Based on this,  we evaluate the role of
 correlations on TMD extractions, comparing 
 forward-rapidity and central-rapidity regions. 

In addition to these bin-to-bin correlations, 
we investigate the impact of other physical 
effects on current studies of 3D proton structure: 
in particular, we concentrate on the role 
of photon emission from the final-state 
DY leptons; on the 
significance of the 
background  from photon-photon scattering 
processes to lepton-pair production; on  
contributions from collinear parton distribution 
functions (PDFs), especially in the kinematic 
region of large longitudinal momentum fractions 
$x$,  relevant to forward production; on 
contributions from finite-order perturbative 
QCD corrections near the boundary of the region used for the 
TMD fits. 

It is worth remarking that vector boson 
production in the forward region, studied 
in this paper from the viewpoint of 
 correlations at low transverse momentum, 
 is also a very sensitive probe of   
small-$x$ dynamics~\cite{Hautmann:2012sh}. 
The investigations of this paper can thus be 
expected to be also relevant for studies of 
transverse hadron structure under 
high-energy evolution (see recent works in~\cite{Caucal:2025mth,Caucal:2025xxh,Caucal:2024bae,Duan:2024qev,Duan:2024qck,Mukherjee:2023snp,Altinoluk:2025ewj,Motyka:2016lta,Taels:2023czt,Li:2026uug,Hautmann:2022xuc}).

\begin{figure}
    \centering
    \includegraphics[width=0.7\linewidth]{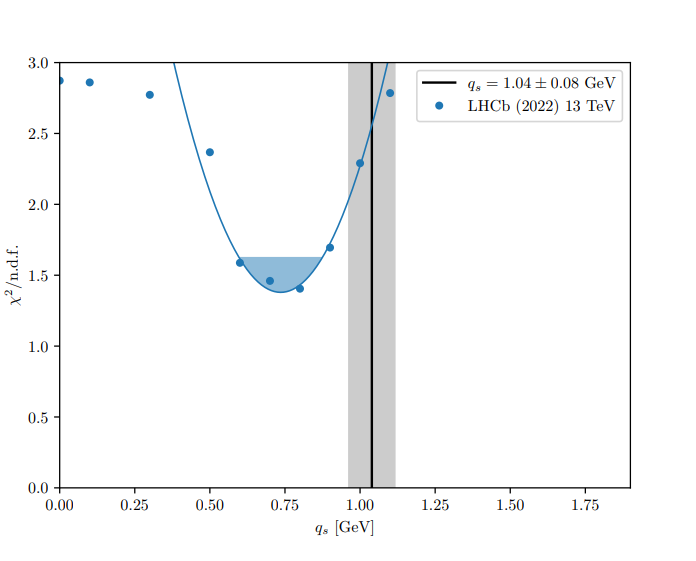}
    \caption{The difference, observed in Ref.~\protect\cite{Bubanja:2023nrd},  between the intrinsic-$k_T$ parameter $q_s$ extracted from the LHCb~\protect\cite{LHCb:2021huf} (blue curve) and CMS~\protect\cite{CMS:2022ubq}  (grey band) DY data at 13\,TeV.}
    \label{fig:LHCbvsCMS}
\end{figure}

To perform our study, we use the same approach 
as Ref.~\cite{Bubanja:2023nrd}, where the intrinsic-$k_T$ parameter was extracted from DY measurements as a function of centre-of-mass energy and lepton-pair invariant mass, using the PB TMD  framework.
A difference in the intrinsic $k_T$ obtained from CMS \cite{CMS:2022ubq} and LHCb \cite{LHCb:2021huf} DY data at 13\,TeV was observed there. The effect is illustrated in Fig.~\ref{fig:LHCbvsCMS} \cite{Bubanja:2023nrd},  where  the $\chi^2/\textrm{n.d.f.}$ (number of degrees of freedom), characterizing  the quality of the LHCb data description,  is shown by the blue curve as a function of the intrinsic-$k_T$ parameter $q_s$ used in the predictions, with the minimum reached for $q_s \approx 0.7$\,GeV. The blue curve is contrasted with the gray band corresponding to the intrinsic-$k_T$ extraction obtained from the CMS data, with $q_s\approx 1.0$\,GeV. 
In the following, we employ the PB framework to investigate the origin of this difference focusing on the data from the LHCb  DY measurement, and identify multiple technical and physical effects that significantly influence the intrinsic-$k_T$ extractions. 

We stress that all such effects  are general, 
and are not by any means limited  
to either the LHCb measurement 
or the PB methodology. 
They involve the statistical treatment 
of DY experimental data---in particular 
correlated uncertainties---and photon contributions 
to lepton-pair final states. These will affect any 
TMD phenomenological study based 
on DY measurements, for any 
centre-of-mass energy, such as for instance 
TMD fits~\cite{Bacchetta:2022awv,Bacchetta:2024qre,Bury:2022czx,Moos:2025sal,Barry:2023qqh,Barry:2025glq,Bubanja:2023nrd,Zhan:2024lym}  and  
studies   
of non-perturbative effects~\cite{Cerutti:2026apy,Zaccheddu:2026pyy,Fernando:2025xzv,Camarda:2025lbt,Aslan:2024nqg}.

The paper is organized as follows. In Sec.~\ref{sec:metho} we 
describe the methodology for the treatment of 
experimental data and theoretical predictions. In 
Sec.~\ref{sec:results} we describe the results of our analysis, 
examining different physical effects,  
and discuss in detail the influence of the transverse 
momentum correlations. 
In Sec.~\ref{sec13} 
we give conclusions. 

\section{Methodology}
\label{sec:metho}

This work investigates the influence of the statistical treatment of experimental 
data for vector-boson production on the extraction of non-perturbative QCD 
parameters, in particular the so-called ``intrinsic $k_T$'' controlling the Gaussian width of TMD parton distributions at the starting mass scale of TMD evolution. 
In general, for a measurement $\vec x$ of dimension $d$ and covariance matrix $C$, the $\chi^2$ metric is defined for a prediction $\vec p$ as 
\begin{equation}
    \label{eq:chi2}
    \chi^2 = \sum_{i,j=1}^d (x_i - p_i) (C^{-1})_{ij} (x_j - p_j).
\end{equation}
In this section, we introduce the measurement $(\vec x, C)$ and prediction $\vec p$ which we will use as a test case in the following. 

The $\chi^2$ definition above takes into account the full covariance matrix $C$.
The diagonal elements of $C$ correspond to the total uncertainty in each bin, while the off-diagonal
elements encode the 
correlations, i.e., how bins are allowed to move with respect to one another. 
Non-zero correlations can arise, for instance, as a result of a 
systematic uncertainty affecting all bins in a similar way.

The best-fit value $\hat q_s$ of $q_s$ is defined as the one minimizing the $\chi^2$.
Although we will not quote explicit uncertainties, we note that the one-sigma 
uncertainty $\Delta\hat q_s$ in the fit result depends on the width of the $\chi^2$
curve.
Indeed, it is usually obtained from a difference of one unit in $\chi^2$ with respect
to the minimum, $\chi^2(\hat q_s\pm\Delta\hat q_s) \equiv \chi^2(\hat q_s) + 1$.
In a complete analysis, such as the one from~\cite{Bubanja:2023nrd}, sources of
uncertainty related to the fitting procedure also need to be accounted for.

Since the CMS measurement~\cite{CMS:2022ubq} provides the breakdown of experimental uncertainties by reporting the 
set of covariance matrices, the analysis~\cite{Bubanja:2023nrd} 
of intrinsic $k_T$  extracted from CMS data 
included the correlation treatment of 
 the systematic and statistical uncertainties. 
 This was however not done in~\cite{Bubanja:2023nrd} when analyzing 
 the LHCb measurement~\cite{LHCb:2021huf} (as well as  
 other centre-of-mass energy measurements): for these cases, 
 only the simplified approach treating all uncertainties as 
 uncorrelated and allowing for a $K$-factor, i.e., normalisation 
 factor, was used in~\cite{Bubanja:2023nrd}. 
 In this paper we revisit the analysis of 
 the LHCb data. First, 
 we take into account all the available information on 
 statistical and efficiency uncertainties, and corresponding correlations,  
 to reconstruct covariance matrices, where 
 possible (see Subsec.~\ref{sec:methodology-data}).  
 Next,  we  investigate  the  impact of normalisation factors and correlations     
 on the $\chi^2/{\textrm{n.d.f.}}$ calculations and, as a consequence, on 
 the intrinsic-$k_T$ extraction (see Sec.~\ref{sec:results}).  
 In doing this, we also study 
 the influence of several other physical effects, including 
 QED contributions, 
 collinear PDFs, NLO matching, and the range and 
 number of bins in transverse momentum.

\subsection{Data}
\label{sec:methodology-data}

Let us consider  the measurement~\cite{LHCb:2021huf} of the transverse momentum $p_\mathrm T$ of DY muon pairs from $Z$-boson decay performed by the LHCb experiment at a 
center-of-mass-energy of 13\,TeV. 
Data points for the single-differential cross section 
$\mathrm d\sigma/\mathrm dp_\mathrm T$ are available from the corresponding HepData record, as well as double-differential results binned in the transverse momentum 
$p_\mathrm T$ and rapidity $y$ of the muon pair, $\mathrm d^2\sigma/\mathrm dp_\mathrm T\,\mathrm d|y|$.  

The HepData record for the measurement~\cite{LHCb:2021huf} 
 provides the uncertainties as a  
 detailed breakdown accompanied with 
 correlation tables for the statistical and efficiency uncertainties, 
 allowing one to reconstruct the corresponding covariance matrices.
No correlation information is provided for the uncertainties in the luminosity, alignment, closure, background, unfolding, or final-state-radiation (FSR) correction.

\begin{table}
    \caption{%
        Sources of experimental uncertainties  
        from~\protect\cite{LHCb:2021huf} used in this paper 
        and of theoretical uncertainties,   
        and their correlation scheme.
    }
    \label{tab:unc-breakdown}
    \centering
    \begin{tabular}{ll}
        \toprule
        Uncertainty & Treatment \\
        \midrule
        Statistical & Correlation matrix from HepData \\
        Efficiency  & Correlation matrix from HepData \\
        Luminosity  & Full correlation \\
        Alignment   & Full correlation \\
        Closure     & Full correlation \\
        Background  & Uncorrelated \\
        Unfolding   & Uncorrelated \\
        FSR         & Uncorrelated \\
        \midrule
        Scale variations & Full correlation \\
        \bottomrule
    \end{tabular}
\end{table}

In the analysis that follows, 
we will explore  $\chi^2$ calculations in scenarios 
with and without correlations. The sources of uncertainties 
considered by the LHCb  measurement~\cite{LHCb:2021huf}  and the 
treatment of correlations considered in the present paper
are summarized in Table~\ref{tab:unc-breakdown}. 
For the scenarios with correlations,   
we will 
consider that  the luminosity, alignment, and closure are fully correlated between bins. The background uncertainty includes different physical processes 
in different bins, whose shape is, in addition, not entirely known. We will thus keep the bins uncorrelated. 
The unfolding uncertainty accounts for a possible bias in the shape of the
distribution introduced by the unfolding procedure, typically towards the
prediction of the Monte Carlo sample used for unfolding.
As a change in shape is not appropriately described by a fully correlated uncertainty, we will keep it   as 
uncorrelated. Likewise, we will keep the FSR uncertainty uncorrelated because FSR strongly affects the shape of the DY $p_\mathrm T$ distribution.

\begin{figure}
    \centering
    \includegraphics[width=0.6\textwidth]{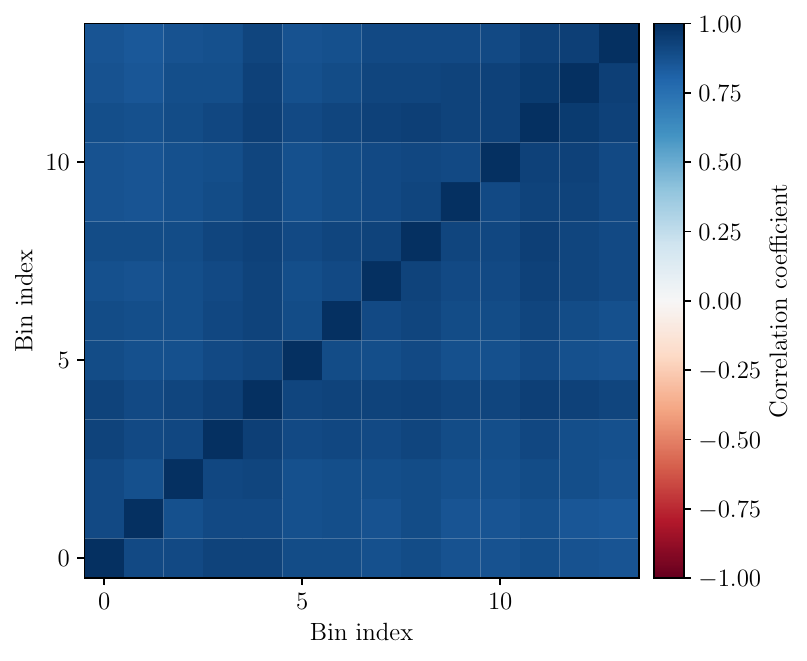}
    \caption{%
        Correlation matrix reconstructed from the LHCb HepData record.
    }
    \label{fig:correlation-matrix}
\end{figure}

 The
resulting correlation matrix is shown in Fig.~\ref{fig:correlation-matrix}.

\subsection{Predictions}
\label{sec:methodology-pred}

The theoretical predictions used for the study carried out in this paper  
follow closely those used in the analysis~\cite{Bubanja:2023nrd}.  
We here limit ourselves to giving a very brief recap of the main elements 
of the calculations, and refer the reader to the 
publication~\cite{Bubanja:2023nrd}, and references therein, for more detail.  

According to the method set out in 
Refs.~\cite{BermudezMartinez:2019anj,BermudezMartinez:2020tys},  predictions for 
the  transverse momentum $p_\mathrm T$ distribution 
of  DY muon pairs  are  obtained as follows.  
Next-to-leading-order (NLO) hard-scattering matrix elements (ME) 
for DY production at 13\,TeV are   
computed from the  
{\scshape MadGraph5\_aMC@NLO}~\cite{Alwall:2014hca}
(hereafter, MCatNLO)  event generator, and are 
matched~\cite{BermudezMartinez:2019anj,Abdulhamid:2021xtt,Yang:2022qgk}
    with  TMD parton distributions which fulfill the 
 parton branching (PB) evolution 
 equations~\cite{Hautmann:2017xtx,Hautmann:2017fcj}  
 and corresponding parton showers,  
 implemented in the  {\sc Cascade}  Monte Carlo event 
 generator~\cite{CASCADE:2021bxe,CASCADE:2010clj}.
 The matching is performed by using the subtractive procedure 
 proposed in Ref.~\cite{BermudezMartinez:2019anj} and further 
 analyzed in Refs.~\cite{Abdulhamid:2021xtt,Yang:2022qgk}:    
 in particular, the
 {\sc Herwig}6~\cite{Corcella:2002jc,Corcella:2000bw} subtraction
 terms are employed in MCatNLO, 
 as they are based on the same angular ordering conditions
 as the PB TMD distributions~\cite{Yang:2022qgk}. 
Final state parton showers
are generated from  {\sc Pythia}6~\cite{Sjostrand:2006za}, including
photon radiation.

It has been shown in 
Refs.~\cite{Martinez:2024mou,Martinez:2024twn} 
that the calculational framework described above 
is able to achieve, in addition to  fixed-order NLO perturbative accuracy, 
the resummation of the 
Sudakov logarithms of $M/p_T$  in the 
$p_T \ll M$ region of the DY spectrum 
(where $M$ is the lepton-pair invariant mass), 
to all orders in $\alpha_s$, 
with next-to-leading-logarithmic accuracy (NLL) 
and, if the physical ``soft-gluon coupling''~\cite{Banfi:2018mcq,Catani:2019rvy}  
is included in the Sudakov evolution, 
with next-to-next-to-leading-logarithmic  
(NNLL) accuracy.  In the  calculations that 
follow, we will limit ourselves to  NLL accuracy, as in 
Ref.~\cite{Bubanja:2023nrd}.  

 Also, it has been shown in 
Refs.~\cite{BermudezMartinez:2021lxz,BermudezMartinez:2022bpj} that the treatment of the high-$p_T$ region in the calculational 
 framework described above 
 can be improved by including higher-order contributions from 
the production of  multiple hard jets via 
multi-jet merging techniques. However, in the calculations that follow 
we will focus on the low $p_T$ region, $p_T \ll M$,  and 
we will not consider the multi-jet merging predictions. 

The PB TMD evolution equations on which our calculational 
framework is based imply a soft-gluon resolution scale $z_M$~\cite{Hautmann:2017xtx}, which may be taken to be  
a fixed~\cite{BermudezMartinez:2018fsv} (i.e., constant) 
or running~\cite{Hautmann:2019biw} (i.e., branching-scale 
dependent) scale.\footnote{A discussion of the 
physical implications 
of fixed or running resolution scales may be found in 
Ref.~\cite{Hautmann:2025fkw}, emphasizing 
the relationship with non-perturbative Sudakov 
contributions and 
Collins-Soper kernel.}  
In the study that follows we will do the same as in 
Ref.~\cite{Bubanja:2023nrd}, and will take fixed $z_M$.   

The evolution equations for the TMD distributions 
have non-perturbative boundary conditions at the 
initial evolution scale $\mu_0$,  given by 
 $ {\cal A}_a  ( x, {\bf k}, \mu_0^2 )$, where 
    $ {\cal A}_a  $ is the TMD distribution of flavor $a$  
as a function of longitudinal momentum fraction $x$ and 
transverse momentum $k_T$, and $\mu_0$ is taken to be 
of order $\mu_0 \sim {\cal O}$(1\,GeV).  In our 
calculational framework, 
 $ {\cal A}_a  ( x, {\bf k}, \mu_0^2 )$ represents 
 the intrinsic   $k_T$  distribution. 
Following 
Ref.~\cite{Bubanja:2023nrd},  
for simplicity in the calculations that follow 
we  parameterize  $  { {\cal A}}_a(x,{\bf k},\mu^2_0)   $ 
in the form 
\begin{equation}
\label{TMD_A0}
  { {\cal A}}_a(x,{\bf k},\mu^2_0)  =  f_{a} (x,\mu_0^2)  
\cdot \exp\left(-| k_T^2 | / 2 \sigma^2\right) / ( 2 \pi \sigma^2) 
\; , 
\end{equation}
with $ | k_T^2 |  = {\bf k}^2 $ and  the width of the Gaussian transverse-momentum 
distribution given by  
$ \sigma  =  q_s / \sqrt{2} $,   
independent of parton flavor  and $x$, where 
$q_s$ is the intrinsic-$k_T$ parameter.  
The dependence on the longitudinal momentum $x$ 
is given by $f_a$, which is 
fitted~\cite{BermudezMartinez:2018fsv} to 
inclusive deep-inelastic scattering data. 
More complex parameterizations, taking into account 
$x$-dependence and flavor dependence of $q_s$, 
are of course possible, for instance following 
the ideas of Ref.~\cite{Bury:2022czx}, which  
 are in fact adopted in 
recent CSS fits~\cite{Bacchetta:2024qre,Moos:2025sal}. 
For the present study, however, we stick to the 
form in Eq.(\ref{TMD_A0}), as our main purpose is to 
investigate the role of correlations and other physical 
and statistical issues, and in this context we prefer 
to keep the parameterization as simple as possible. 
We plan to return to the issues of 
$x$-dependence and flavor dependence in future studies. 

  To explore the intrinsic-$k_T$   parameter space with 
  our PB TMD predictions, 
 following the approach of Ref.~\cite{Bubanja:2023nrd}  
 we generate different TMD template samples which provide 
  replicas of the 
  TMD parton distributions~\cite{BermudezMartinez:2018fsv} 
  characterized by different values of the  intrinsic-$k_T$   parameter 
  $q_s$.  We start with the TMD set 
PB-NLO-HERAI+II-2018-set2  (hereafter, PBSet2), 
  corresponding to $q_s=\SI{0.5}{\giga\electronvolt}$; then we generate replicas,
scanning $q_s$ in steps of \SI{0.1}{\giga\electronvolt}.
For each TMD replica, we perform the 
matching, as described above, 
to obtain the MCatNLO + {\sc Cascade} Monte Carlo prediction for 
the $p_\mathrm T$ distribution of  DY muon pairs. 
Each prediction, obtained using the {\sc Rivet}~\cite{Buckley:2010ar}
analysis tool, is then compared to the experimental measurement.
 The $\chi^2/\rm{n.d.f.}$ is calculated to test the fit quality. 
 The details of the $\chi^2/\rm{n.d.f.}$ calculations, and 
 the results that follow, will be 
 described in the next section. 
 For each of the $\chi^2/\rm{n.d.f.}$ setups, $\hat q_s$ 
 will be obtained from a plot of $\chi^2/\rm{n.d.f.}$ as 
 a function of $q_s$.  

In addition to the experimental uncertainties, we include the statistical and
scale uncertainty in the prediction in the $\chi^2$ calculation.
The latter is obtained by varying the factorization and renormalization scales 
by factors of two and one half in the collinear PDF and ME  calculation. 
We calculate the uncertainty from the largest deviation from the central value,
excluding the cases where the two scales vary in opposite directions.
In the computations that follow, we assume the uncertainty to be  fully correlated across bins, that is, we use the assumption that  the scales are 
not  allowed to vary from bin to bin.
Since it is roughly flat in $p_\mathrm T$, the scale uncertainty is comparable
to an uncertainty in the normalization.
The uncertainties in the collinear part of the TMD are not used because they
are only available for $q_s=\SI{0.5}{\giga\electronvolt}$, and evaluating them
for other points would be very demanding computationally.
A test based on the available uncertainties at $q_s=\SI{0.5}{\giga\electronvolt}$
indicated a slight improvement in $\chi^2$ without changing $\hat q_s$.

\begin{figure}
    \centering
    \includegraphics[width=0.75\linewidth]{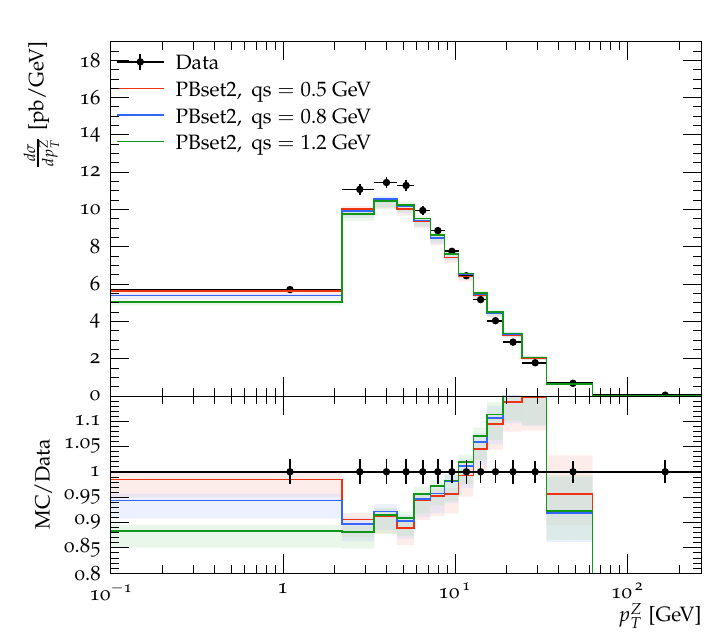}
    \caption{DY predictions obtained with three different replicas of PB-NLO-HERAI+II-2018-set2~\protect\cite{BermudezMartinez:2018fsv} (PBSet2) differing by the intrinsic-kt parameter $q_s$, compared with the LHCb measurements~\protect\cite{LHCb:2021huf}.}
    \label{fig:defaultPBset2fullrange}
\end{figure}

To start with, DY $p_T$ predictions compared to the LHCb DY data at $13$\,TeV, obtained with three replicas of PBSet2 with three different values of the intrinsic-$k_T$ parameter $q_s$, are shown in Fig.~\ref{fig:defaultPBset2fullrange}. As a purely non-perturbative effect, the intrinsic-$k_T$  plays a role in the first bin of the spectrum, for $p_T<2$\,GeV. 
As the measurement is performed at the scale of the $Z$-boson 
mass, 
however, the intrinsic-$k_T$  extractions are also sensitive to the higher $p_T$ range of the transverse momentum of the DY pair considered in the analysis and to the importance given to each bin. 

\section{Results and discussion} 
\label{sec:results}

\begin{figure}
    \centering
    \includegraphics[width=0.75\linewidth]{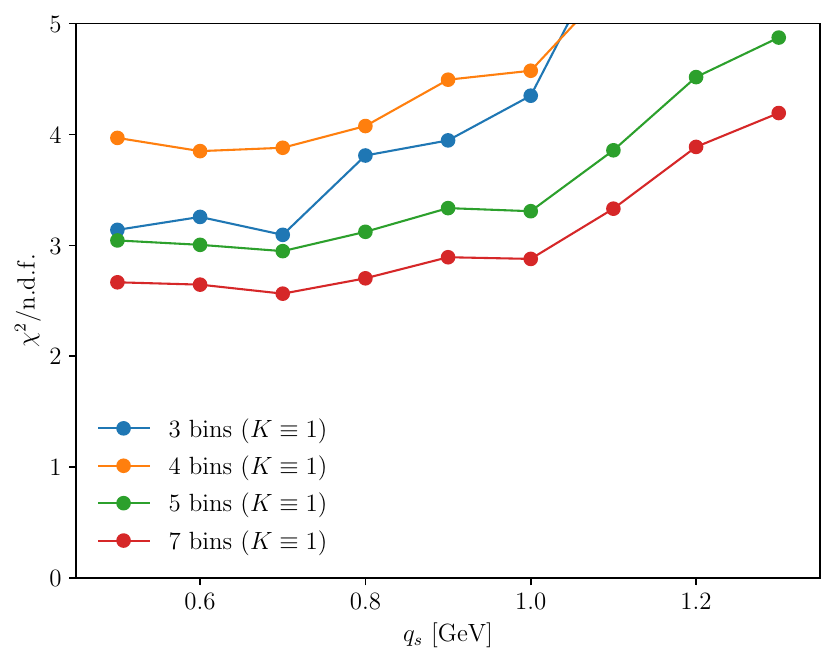}\\
    \includegraphics[width=0.75\linewidth]{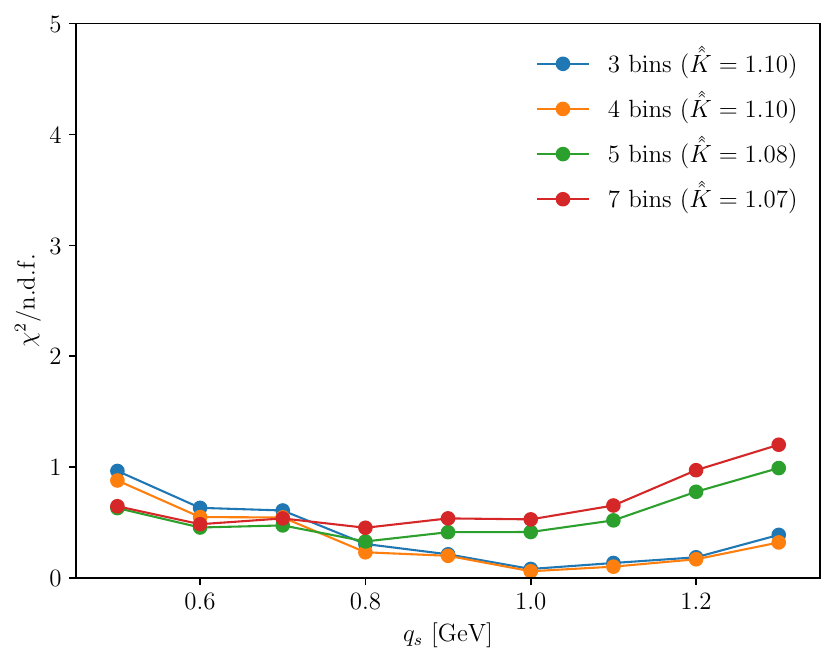}
    \caption{%
        $\chi^2/\rm{n.d.f.}$ vs $q_s$ for different numbers of bins, ignoring correlations
        (diagonal covariance matrix).
        Results are shown with the $K$-factor set to 1 (upper panel) and with a profiled
        $K$-factor (lower panel).
    }
    \label{fig:NoCorrelation}
\end{figure}

As a  first step of this analysis, we consider the 
computation of the $\chi^2/\rm{n.d.f.}$ in Eq.~(\ref{eq:chi2}) 
without taking into account any correlations and without introducing any normalisation factor  to rescale the prediction. 
The result can be seen in the upper panel of Fig.~\ref{fig:NoCorrelation}, for different numbers of bins. One can see that a wide minimum is obtained, with a stable position around $q_s=0.6$\,GeV with respect to the change of the number of bins, if the number of bins is between $4$ and $7$. The value of the best $\chi^2/\rm{n.d.f.}$ itself is  between $2$ and $4$. One can see from Fig.~\ref{fig:defaultPBset2fullrange} that most of the bins in the $p_T$ range 2--10\,GeV are not really well described. 

In the next two subsections, we describe the next steps of 
our analysis, in which we motivate and introduce the normalisation procedure, and investigate correlations. We then 
examine  physical effects---including 
 collinear PDFs, final state photon emission, photon-photon background, details of the perturbative matching---which 
 influence the comparison of 
theory predictions with experimental data,  
and present results for the extraction of nonperturbative 
TMD parameters.

\subsection{Normalisation procedure}
\label{subsec:norm}

The normalisation of the prediction affects, of course, the $\chi^2/\rm{n.d.f}$. 
A distribution with the correct shape but incorrect normalisation will often get a worse $\chi^2$
than distribution with the wrong shape but better normalisation.
This can be seen in the low-$p_\mathrm T$ part of Fig.~\ref{fig:defaultPBset2fullrange}: the red
curve comes much closer to data in the first bin than the green curve, but the latter describes the
overall shape better.
In our case, we do not expect our NLO calculation, based on TMDs whose collinear part was extracted
from HERA data, to predict the  integrated cross section accurately in the LHCb phase space.
The effect we are interested in---intrinsic $k_\mathrm T$---mainly affects the shape of the 
distribution.
This motivates the use of a $K$-factor to rescale the prediction by a constant value. 
Since the $K$-factor affects the $\chi^2$ in a non-trivial way, the procedure used to determine its
value can affect the fit results.
The $K$-factor also adds one free parameter to the fit, reducing the number of degrees of
freedom by one.

Several approaches are possible to calculate $K$.
A common solution is to use a ratio of integrated cross sections:
\begin{equation}\label{eq:k0}
    K_0 = \frac{\sum_{b\in\text{bins}} \text{data (bin $b$)}}{\sum_{b\in\text{bins}} \text{prediction (bin $b$)}}.
\end{equation}
Here, the sums run over some set of bins  taken into consideration for the normalisation; this 
can be the complete distribution or only part of it.
It is clear from Fig.~\ref{fig:defaultPBset2fullrange} that one can get very different results  
depending on which bins are considered: using the first 4 bins would yield $K_0\approx1.1$ while
using the last 4 bins would yield $K_0\approx0.85$.
Another drawback of this formula is that it ignores uncertainties: a very uncertain bin has the
same weight as one that is known very precisely.
Finally, $K_0$ itself should carry an uncertainty, which is usually ignored.

In this work, we use a statistically motivated method by \emph{profiling} the normalisation 
parameter $K$.
For each value of $q_s$, we determine the $K$-factor $\hat K(q_s)$ that minimizes the
$\chi^2$ from Eq.~\eqref{eq:chi2},
\begin{equation}\label{eq:khat}
    \hat K(q_s) 
    =
    \arg\min_{K}\left(
        \sum_{i,j=1}^d (x_i - K p_i) (C^{-1})_{ij} (x_j - K p_j)
    \right),
\end{equation}
where the dependence on $q_s$ enters through both $p$ and $C$.
In keeping with statistics literature, we denote by $\doublehat K$ the value of $\hat K$ at the 
minimum, $\doublehat K \equiv \hat K(\hat q_s)$.

The construction above yields the best possible $K$-factor for each value of $q_s$ according to
the same metric used to extract $q_s$.
In contrast to Eq.~\eqref{eq:k0}, the bins used to calculate the $\chi^2$ in the profiling are
the same as the one used to extract $q_s$.
As we will see in the next paragraph, this is a maximally conservative choice.
It is also obvious that the determination of $\hat K(q_s)$ takes uncertainties into account,
since the covariance matrix is included in the minimized $\chi^2$.

Profiling a parameter usually leads to an increase in the uncertainty extracted from the
fit, because minimizing the $\chi^2$ for each value of $q_s$ tends to broaden the curve 
around the minimum, pushing the crossing of the $\Delta\chi^2=1$ threshold farther away.
This may be understood as the contribution of the normalization to the uncertainty on 
the extracted value of $q_s$.
Using the same bins for calculating the $\chi^2$ and $\hat K(q_s)$ maximizes this effect, 
leading to a larger uncertainty.

Results obtained with $K$-factor are shown in the lower panel of Fig.~\ref{fig:NoCorrelation} for
different numbers of bins,
with the best-fit $K$-factors $\doublehat K$ written in the legend.
Both the $\chi^2$ calculation and the $K$-factor still use diagonal covariance matrices.
Profiling results in a widening of the $\chi^2(q_s)$ curve around its minimum, increasing the width of the
confidence interval in $q_s$.
This corresponds to adding an uncertainty for the unknown $K$-factor.
One can see that using the $K$-factor improves  $\chi^2/\rm{n.d.f}$  compared to the results shown in the upper panel of Fig.~\ref{fig:NoCorrelation}, and that the minima are shifted towards higher values, with the exact value of the best $\chi^2/\rm{n.d.f}$ being number-of-bins dependent. The result obtained for 5 bins is very similar to the treatment of LHCb data in \cite{Bubanja:2023nrd}, reproducing the outcome of that work\footnote{In \cite{Bubanja:2023nrd}, no profiling was used to obtain the $K$-factor: the $K$-factor was obtained from Eq.~\eqref{eq:k0}. We checked that this choice does not affect the results from~\cite{Bubanja:2023nrd} significantly.}.

\subsection{Correlations}
\label{subsec:corr}

We have so far been using diagonal covariance matrices. This is an approximation which is commonly used
by computational tools in high-energy physics that lack support for propagating the full covariance
matrix.
In this case, Eq.~\eqref{eq:chi2} reduces to the well-known form in which the
differences between bins are divided by their respective uncertainties.
Neglecting the off-diagonal elements of the covariance matrix gives more freedom to the
fit, especially for systematics-dominated measurements like the
$Z$-boson  $p_\mathrm T$, resulting in an artificially better agreement between data and predictions.
This also significantly reduces the constraining power of the data, resulting in larger uncertainties
for the extracted parameters.

\begin{figure}
    \centering
    \includegraphics[width=0.75\linewidth]{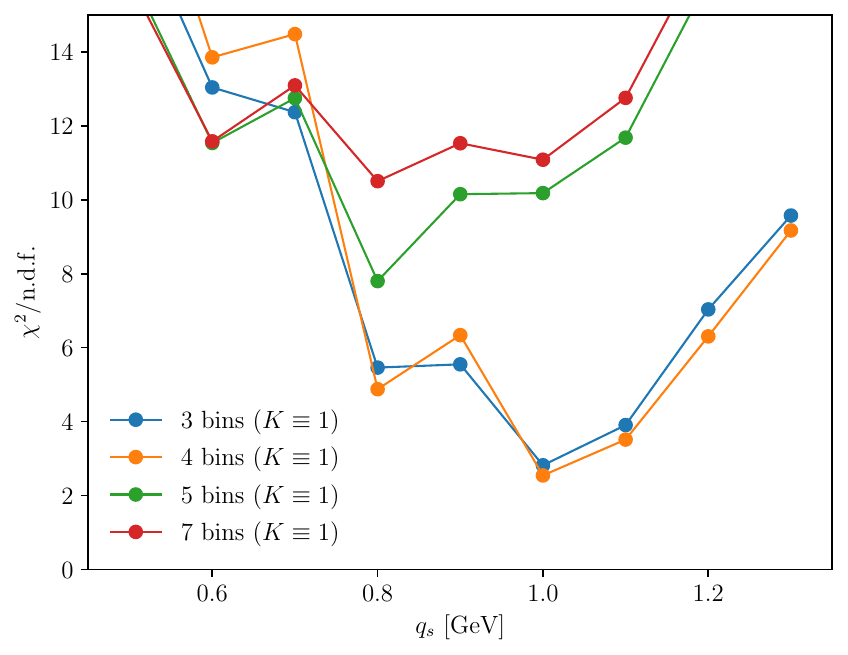}
    \includegraphics[width=0.75\linewidth]{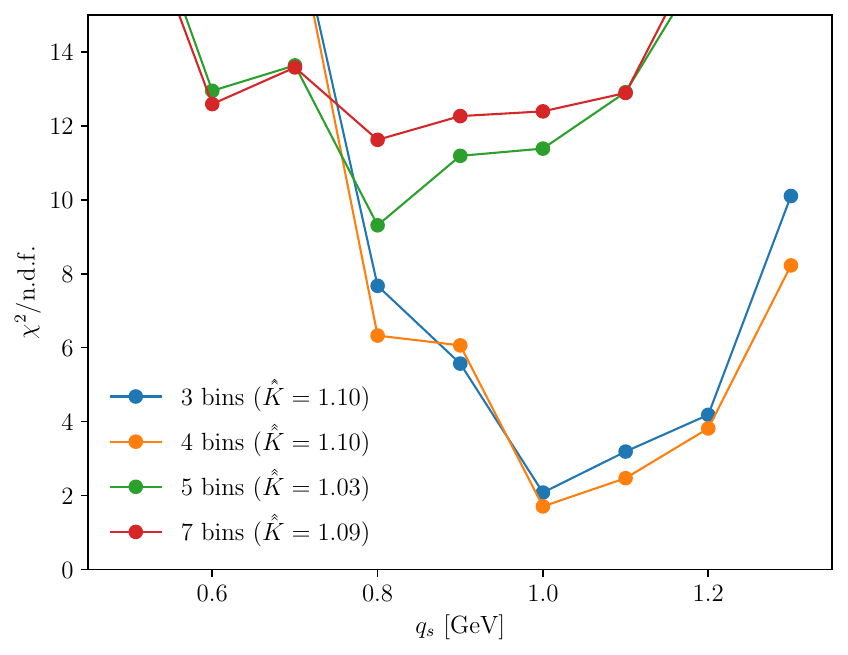}
    \caption{%
        $\chi^2/\rm{n.d.f.}$ vs $q_s$ for different numbers of bins, accounting for correlations
        (using the complete covariance matrix).
        Results are shown with the $K$-factor set to 1 (upper panel) and with a profiled
        $K$-factor (lower panel).
    }
    \label{fig:Correlation}
\end{figure}

To demonstrate this effect, we show in Fig.~\ref{fig:Correlation} the $\chi^2$ curves
obtained with correlations for different numbers of bins.
As in Fig.~\ref{fig:NoCorrelation}, the normalization given by the prediction is used
for the upper panel ($K\equiv1$).
Compared to the upper panel of Fig.~\ref{fig:NoCorrelation}, narrower minima are obtained, with
the position of the minima shifted towards higher $q_s$ values, dependent on the number
of bins included in the calculation. 
The values of the best $\chi^2/\rm{n.d.f}$ are very high with five or seven bins, but 
decrease for three or four bins.

The lower panel of Fig.~\ref{fig:Correlation} uses a profiled $K$-factor, whose best-fit 
values $\doublehat K$ differ from the ones in Fig.~\ref{fig:NoCorrelation} since
correlations also enter Eq.~\eqref{eq:khat}.
A similar method of $\chi^2$ calculation was used in \cite{Bubanja:2023nrd} to make the intrinsic-$k_T$ determination from \SI{13}{\tera\electronvolt} CMS DY measurement\footnote{Again, in \cite{Bubanja:2023nrd}, no profiling was used to obtain the $K$-factor:  the $K$-factor was obtained from Eq.~\eqref{eq:k0}, but the results are unaffected.}.
One can see that,  by including both $K$-factor and correlations, narrow minima  with good $\chi^2/\rm{n.d.f.}$ are obtained with three or four bins.
Minimas are shifted to the right compared to Fig.~\ref{fig:NoCorrelation}, and the exact value of the best-fit  $q_s$ is number-of-bins dependent. 
The $q_s$ value corresponding to the best $\chi^2/\rm{n.d.f.}$ is close to \SI{1.0}{\giga\electronvolt}.
The curve for $4$ bins from the lower panel of Fig.~\ref{fig:Correlation} is used further on in figures labeled as the ``Reference''.

\begin{figure}
    \centering
        \includegraphics[width=0.75\linewidth]{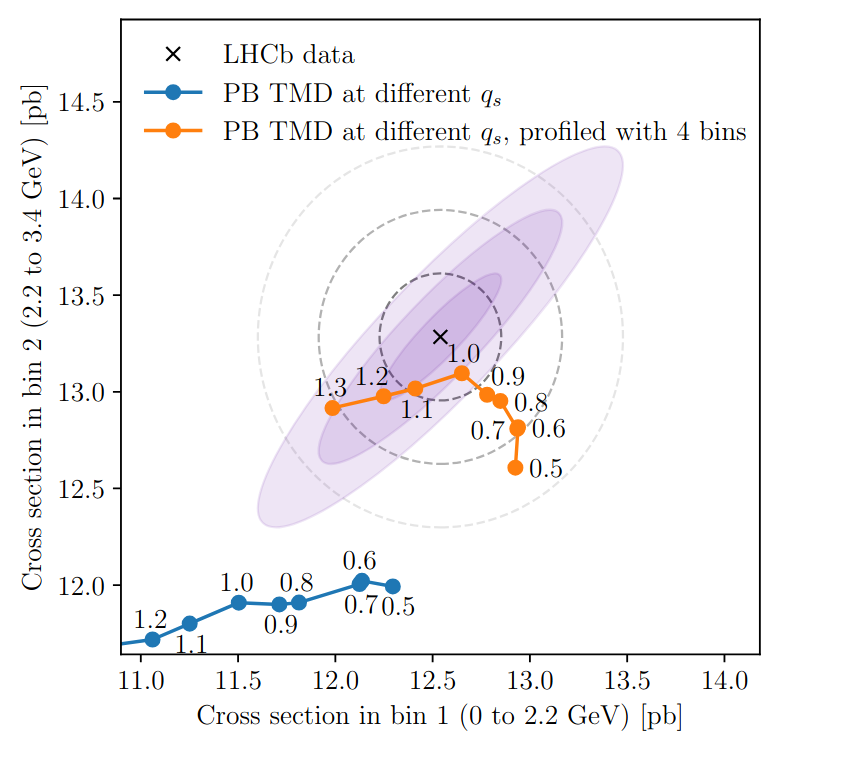}
    \caption{
        The effect of including the correlations: the LHCb measurement in the first two $p_\mathrm T$ bins (black cross) is
        compared to the PB TMD prediction  without $K$-factor (blue) and with $K$-factor (orange).
        The uncertainty in the measurement with correlations is shown with shaded ellipses at $\Delta\chi^2=1$, 4, and 9, corresponding to
        one, two, and three standard deviations respectively.
        The uncertainties in the prediction depend on $q_s$ and are hence not included in the ellipses.
        The dashed circles correspond to the 
        uncertainties in the measurement without
        correlations (again, one, two and three standard deviations). 
    }
    \label{fig:Corr-noCorr-Elipses}
\end{figure}

The effect of using the correlations is illustrated in Fig.~\ref{fig:Corr-noCorr-Elipses}.
The horizontal and vertical axes correspond to the first and second bins, respectively.
The cross at the center of the plot represents the cross-sections in the  first two bins of the $p_\mathrm T$
spectrum measured by LHCb.
The uncertainty in the measurement, taking correlations into account, is shown with
diagonal ellipses corresponding to $\chi^2$ differences of one, four, and nine.
The same uncertainty, but ignoring the effect of correlations, is shown as dashed circles.
Finally, the PB TMD prediction obtained with (orange) and without (blue) $K$-factor for different values of $q_s$ is shown;
uncertainties in the prediction are not represented.

All PB predictions without $K$-factor (blue)  are far away from the data.
When neglecting correlations, the point at $q_s = 0.5\,$GeV is closest to the data, indicating that a
small value of $q_s$ will be favored.
When taking correlations into account, however, a larger value of $q_s \approx 1\,$GeV matches the
data more closely.
In this 2D projection, values between 1.1 and 1.2 are within the $1\sigma$ ellipse.
Moreover, the confidence interval is significantly reduced when taking correlations into account, as a function of the angle between the PB line and the major axis of the ellipses.
With the profiled $K$-factor (orange), the predictions move closer to the data across diagonal lines.
The resulting $\chi^2$ reduction is visible in Fig.~\ref{fig:Correlation}, although the loss of one
degree of freedom means that the $\chi^2/\rm{n.d.f.}$ does not always improve.

\subsection{Collinear PDF}
\begin{figure}
    \centering
        \includegraphics[width=0.75\linewidth]{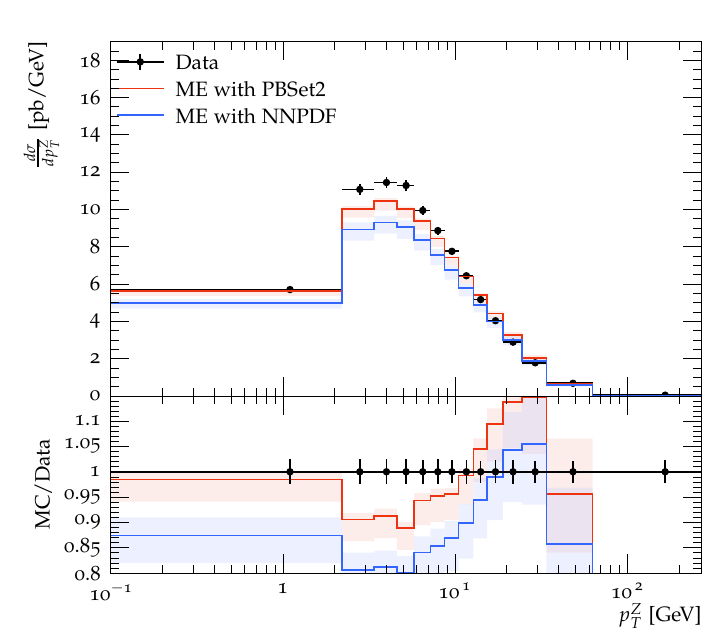}
    \caption{The effect of using different collinear PDFs in the ME generation on the DY $p_T$ spectrum. }
    \label{fig:MEPBset2vsNNPDF}
\end{figure}

\begin{figure}
    \centering
    \includegraphics[width=0.75\linewidth]{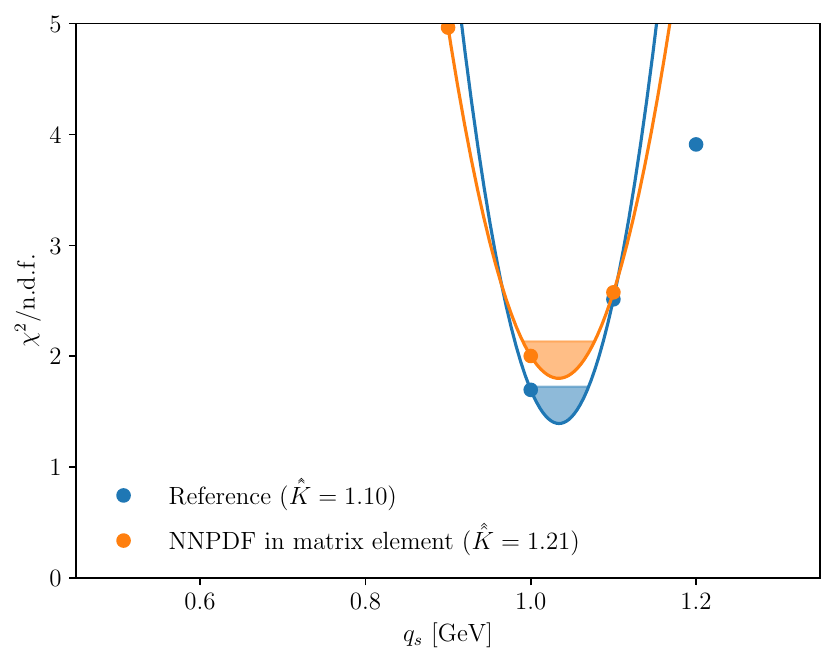}
    \caption{%
        $\chi^2/\rm{ndf}$ vs $q_s$ for different collinear PDF, calculated with $K$-factor and correlations.
    }
    \label{fig:KfactorWithCorrelation-NNPDF}
\end{figure}

\begin{figure}
    \centering
    \includegraphics[width=0.75\linewidth]{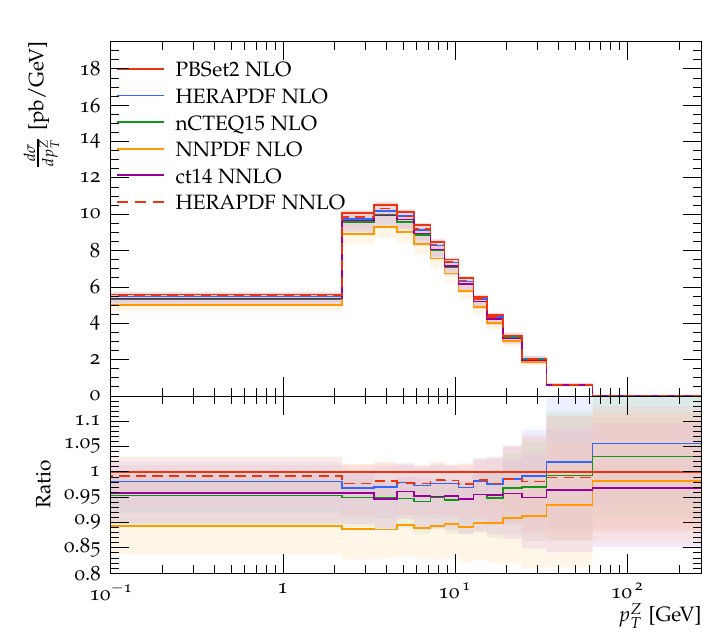}
    \caption{%
     The prediction for DY  $p_{T}$ obtained with MEs generated with different collinear PDFs indicated by the legend, matched with PBset2 TMD. 
    }
    \label{fig:collPDFS}
\end{figure}

The results for the TMD determination and extraction of the 
intrinsic-$k_T$ parameter $q_s$ can be potentially affected 
by the PDF used for ME generation, in a manner 
 similar to the ``PDF bias'' effect pointed out in Ref.~\cite{Bury:2022czx}. This is 
studied in Figs.~\ref{fig:MEPBset2vsNNPDF}--\ref{fig:collPDFS}. 

In Fig.~\ref{fig:MEPBset2vsNNPDF},   two predictions for the 13\,TeV DY $p_T$ spectrum are shown, obtained with NLO MEs generated with 
two different PDF sets, the integrated-TMD 
PBSet2~\cite{BermudezMartinez:2018fsv} and the NNPDF30$\_$nlo$\_$as$\_$0118 (NNPDF)~\cite{NNPDF:2014otw} sets.  In   both cases,  the matching  with the default PBSet2 TMD 
distributions~\cite{BermudezMartinez:2018fsv,BermudezMartinez:2019anj} (that is, assuming $q_s=0.5$\,GeV) is performed, and the results are compared with the LHCb data. One can see that the normalisations obtained from the two predictions differ substantially.

In Fig.~\ref{fig:KfactorWithCorrelation-NNPDF}, the impact of using those two different collinear PDFs in ME generation 
on the intrinsic-$k_T$ extraction, upon including $K$-factor and correlations, is shown. We see that,  
for the two considered PDFs, there is no large 
effect in the best-fit $q_s$ or in the quality of the agreement.
That  is, although the predictions for the DY $p_T$ 
in Fig.~\ref{fig:MEPBset2vsNNPDF} differ, since 
 the collinear PDF  mainly affects the overall normalisation of the simulation, our TMD extraction 
 based on  the normalisation procedure 
  in Sec.~\ref{subsec:norm} and 
  correlations in Sec.~\ref{subsec:corr}
is able to reabsorb most of the collinear-PDF effects 
and is thus robust with respect to PDF variations.

This observation can be further verified by checking the DY $p_{T}$  predictions at $13$\,TeV obtained with MCatNLO MEs generated with other collinear PDFs, two NLO ones, nCTEQ15 \cite{Kovarik:2015cma}, and HERAPDF20$\_$NLO$\_$EIG~\cite{H1:2015ubc}, and two NNLO ones, CT14MC2nnlo \cite{Hou:2016sho}, and HERAPDF15NNLO  EIG \cite{H1:2015ubc}. All of the MEs are matched with the PBSet2 TMD. The resulting predictions are illustrated 
in Fig.~\ref{fig:collPDFS}. It can be seen that at low and middle $p_{T} < 10$\,GeV the difference is, mostly, a constant shift. 
A non-negligible difference in shape can however be observed 
for the HERAPDF sets, both NLO and NNLO, in the first $p_T$ bin,  that is, in the region which is most sensitive to the intrinsic-$k_{T}$ extractions. 

Overall, we regard the results in Figs.~\ref{fig:KfactorWithCorrelation-NNPDF} 
and~\ref{fig:collPDFS} as indicating that 
our TMD extraction methodology is sufficiently stable 
with respect to collinear-PDF effects. We leave to future 
work the investigation of the potential impact of 
differences in shape at the lowest $p_T$ such as those 
noted above for the HERAPDF sets. These may become 
significant, with increasing accuracy,  
particularly  for the analysis of the 
forward rapidity region probed by LHCb, 
given its sensitivity to the effects of 
large-$x$ PDFs~\cite{Cole:2026eex,Fu:2023rrs,Fiaschi:2022wgl}.

\subsection{Final state radiation}
\label{Sec:QEDcorr}

\begin{figure}
    \centering
    \includegraphics[width=0.75\linewidth]{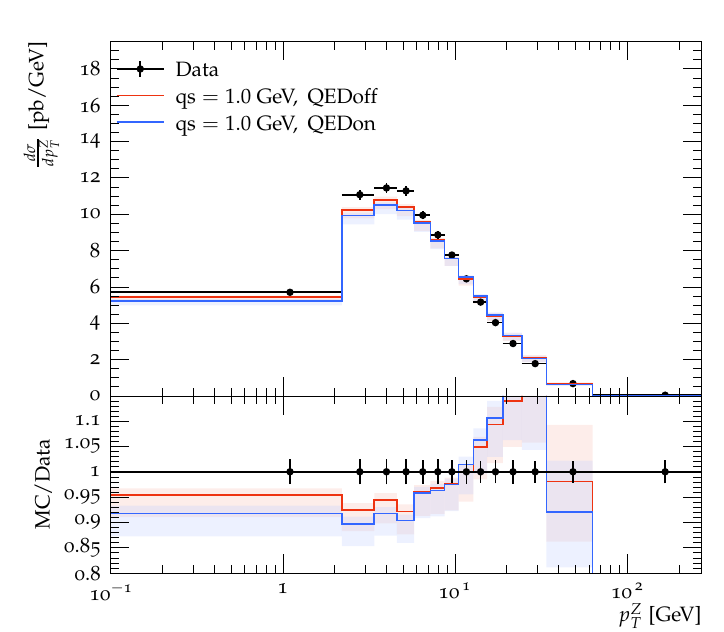}
    \caption{
       DY predictions
        with and without QED correction (obtained by switching on and off the final state radiation), compared to the LHCb data.}
    \label{fig:predictions-QEDonOff}
\end{figure}

It was observed in \cite{Frederix:2018nkq,CMS:2022ubq,Bubanja:2023nrd} that photons emitted from the final-state
leptons reduce the mass of the lepton pair and shift the visible $p_\mathrm T(\mu\mu)$ distribution.
This is also illustrated in Fig.~\ref{fig:predictions-QEDonOff} where the DY predictions with final state radiation turned on and off are compared to LHCb data. 
When analyzing the data, this effect can or not be corrected for, resulting in different phase space definitions.
The LHCb and CMS measurements~\cite{LHCb:2021huf,CMS:2022ubq} differ in this regard.
The Monte Carlo generation of the prediction should be set up according to the data set to be described.

\begin{figure}
    \centering
    \includegraphics[width=0.75\linewidth]{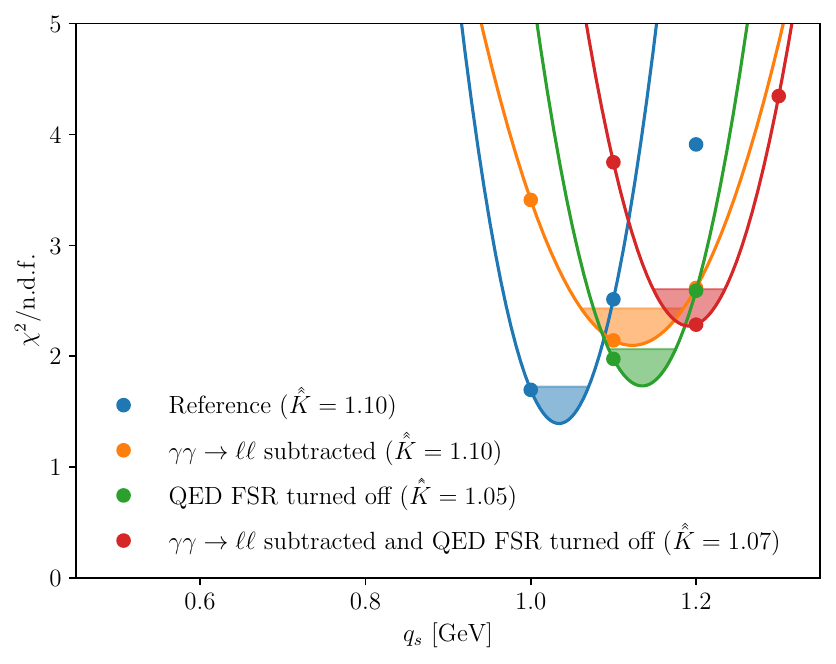}
    \caption{
        $\chi^2/\rm{ndf}$ vs $q_s$, calculated with $K$-factor and correlations. Predictions are shown 
        with and without QED correction in the final state, and with and without subtracting the 
        photon-induced background.
    }
    \label{fig:KfactorWithCorrelation-QEDonOff}
\end{figure}

The LHCb measurement is provided at the Born level, i.e., it was corrected for the
effect of final-state radiation.
This is unlike the {\sc Rivet}~\cite{Buckley:2010ar} routine LHCB$\_$2021$\_$I$1990313$, which is based on 
muons dressed with photons within $\Delta R<0.1$.
To more closely match this phase space, we turn off QED radiation by switching off the final-state
shower, enabling the use of the {\sc Rivet} routine without changes.
The effect of this change, keeping all other elements of the calculation as in the 
4-bin curve from the lower panel of Fig.~\ref{fig:Correlation} (``Reference''),
can be seen in Fig.~\ref{fig:KfactorWithCorrelation-QEDonOff}.
Turning off QED FSR leads to a softer $p_\mathrm T$ spectrum, which is compensated 
by an increase of the best-fit $q_s$ by about 100\,MeV.
The fit quality is not significantly affected.

\subsection{Photon-induced background}

\begin{figure}
    \centering
    \includegraphics[width=0.6\linewidth]{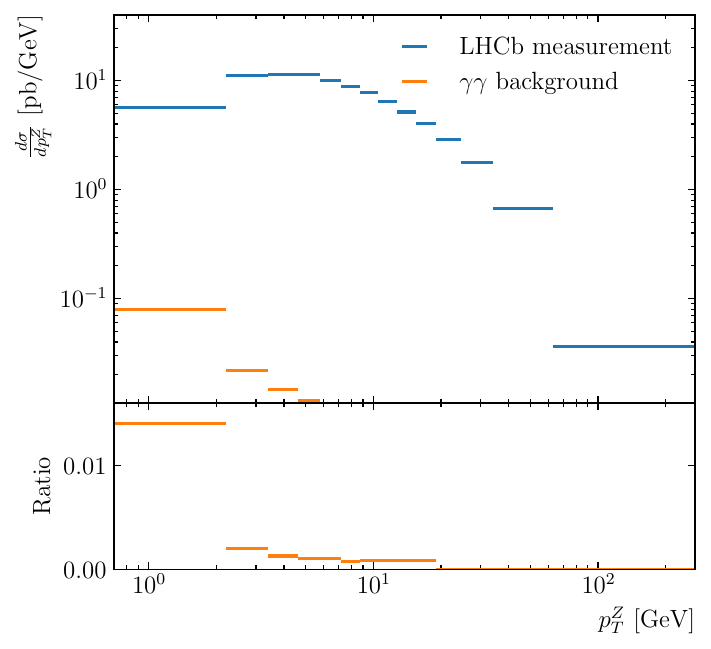}
    \caption{%
        Comparison of the total photon-induced background predicted by LPAIR and the cross
        section measurement of LHCb.
        Uncertainties are not shown.
    }
    \label{fig:GammaGamma}
\end{figure}

The LHCb measurement of DY $p_\mathrm T$ does not consider photon-induced muon pair production,
$\gamma\gamma\to\mu\mu$.
Three cases can be separated for this process depending on whether the photons are produced
elastically by the protons: elastic-elastic, elastic-inelastic, and inelastic-inelastic.
While the total cross sections are small in the considered $m_{\mu\mu}$ range, the events are
concentrated at low $p_\mathrm T$, especially in the fully elastic case.
We simulate it using LPAIR~\cite{Vermaseren:1982cz, Baranov:1991yq} interfaced with {\sc Pythia}6~\cite{Sjostrand:2006za},
using the default QED-PDF of Suri–Yennie~\cite{Suri:1971yx}.
The predicted cross section is shown in Fig.~\ref{fig:GammaGamma}.
The contribution is about 1.5\% in the first bin, which can affect the extracted value of $q_s$.
We assign a 10\% uncertainty to the predicted cross section, fully correlated across $p_\mathrm T$ bins.
It has been checked in~\cite{CMS:2022ubq} that using the more recent QED-PDF, LUXqed-17~\cite{Manohar:2016nzj},
modifies the cross section by less than 6\%, which is covered by this uncertainty.

In Fig.~\ref{fig:KfactorWithCorrelation-QEDonOff}, the effect of subtracting the photon-induced background is shown.
This decreases the cross section in the first bin by about 1.5\%, changing the shape of
the $p_\mathrm T$ spectrum.
Indeed, we see that the best-fit value of $q_s$ shifts by about 0.06\,GeV
to the right to compensate,  which is a large shift compared to the experimental uncertainties.

The QED effects, both from the final state radiation and photon-induced background, were included in \cite{Bubanja:2023nrd} to analyse the CMS data. The result shown in red in Fig.~\ref{fig:KfactorWithCorrelation-QEDonOff} is then the closest to the treatment of CMS in \cite{Bubanja:2023nrd}. Given that, we observe a difference between the best $q_s$ value extracted from LHCb ($q_s \approx1.2$\,GeV) and CMS ($q_s\approx 1.0$\,GeV).

\subsection{Comments}
\label{sec:comm}

We are now in a position to provide 
a few general comments on the findings 
of our study, referring back to 
Fig.~\ref{fig:LHCbvsCMS} obtained 
from the analysis in 
Ref.~\cite{Bubanja:2023nrd}. 
By comparing 
Fig.~\ref{fig:LHCbvsCMS} and 
Fig.~\ref{fig:KfactorWithCorrelation-QEDonOff}, 
we see that the full treatment of the 
correlations in the 
LHCb measurements~\cite{LHCb:2021huf} 
and of the normalisation factor $K$, 
along with the appropriate treatment 
of  the QED effects, influences 
the extraction of 
the TMD non-perturbative parameter 
$q_s$  significantly. 
This is a general conclusion, which 
will apply to any phenomenological 
study of TMDs from DY measurements, 
including global fits~\cite{Bacchetta:2024qre,Moos:2025sal,Barry:2025glq} 
and non-perturbative 
parameterizations~\cite{Camarda:2025lbt,Aslan:2024nqg}. 
In all such studies, the statistical 
treatment of transverse momentum 
measurements, including bin-to-bin 
correlations, will influence TMD 
determinations, as will the treatment of 
QED corrections. 

As a result of the analysis leading to 
Fig.~\ref{fig:KfactorWithCorrelation-QEDonOff}, 
the difference in the extractions of $q_s$ 
from the LHCb data set (in this paper) and 
from the 
CMS data set (in Ref.~\cite{Bubanja:2023nrd}) 
is less than 2$\sigma$. As stressed earlier, 
the effects which we have taken into account 
in the present analysis (correlations, $K$-factor, QED contributions) all have a non-negligible 
impact on the $q_s$ extraction.  
At the level of the 2$\sigma$ difference 
observed above, we note further that 
other effects may also be 
 significant, and should thus be taken into account 
in future investigations. One such effect concerns 
the TMD uncertainties. Although 
TMD uncertainties are provided in the 
PB TMD set~\cite{BermudezMartinez:2018fsv} 
with $q_s = 0.5$\,GeV, 
 they are not available for the 
$q_s$ replicas~\cite{Bubanja:2023nrd}  
used in the present study. Thus, we have  
limited ourselves to including perturbative 
scale uncertainties, as described in 
Sec.~\ref{sec:methodology-pred}, in 
the theoretical predictions employed 
for the $\chi^2$ calculations in 
Fig.~\ref{fig:KfactorWithCorrelation-QEDonOff}. 
We leave to future work 
the inclusion of TMD uncertainties in 
addition to the  
perturbative  scale uncertainties, 
and the investigation of how this 
may influence the error bands 
and the positions of the minima in 
Fig.~\ref{fig:KfactorWithCorrelation-QEDonOff}.
We further note that a complete compatibility
test between the results extracted from LHCb
and CMS data should be based on a combined
fit, keeping parameters such as TMD uncertainties
and the normalisation $K$ factors consistent between
the two measurements.

The remarks in the above paragraph 
underline the relevance of a 
complete treatment 
of theoretical systematic uncertainties in 
TMD phenomenology. This is in fact 
 becoming critical for  
determinations of both 
collinear and TMD distributions, see,  
e.g., Ref.~\cite{Bertone:2024snr}, and 
references therein. 

 Another comment concerns  the 
comparison of the 
value for the intrinsic-$k_T$ parameter 
$q_s$ from the PB TMD analysis 
of the LHCb measurements at 
$\sqrt{s} = 13$\,TeV 
performed in this paper 
with the 
intrinsic-$k_T$ extraction of 
Ref.~\cite{CMS:2024goo}, based on  
 the tuning of parton-shower Monte 
Carlo event generators to experimental
DY data.  From 
{\sc Pythia}~\cite{Sjostrand:2014zea} 
Monte Carlo tuning, for example, 
Ref.~\cite{CMS:2024goo} 
obtains intrinsic-$k_T$ values of about 
3\,GeV  at  $\sqrt{s} = 13$\,TeV. 
The result~\cite{CMS:2024goo}  
from Monte Carlo tuning 
 corresponds to a steep rise with 
energy of the intrinsic $k_T$, leading to 
approximately a factor of~3 increase 
at LHC energies compared to 
low-energy experiments at a few tens of~GeV. 
In contrast, our result in 
Fig.~\ref{fig:KfactorWithCorrelation-QEDonOff} 
is near 1\,GeV. This, together with  
the results from the analogous analysis of the 
CMS measurements at $\sqrt{s} = 13$\,TeV 
performed in Ref.~\cite{Bubanja:2023nrd} and 
from the analysis of low-energy experimental 
data also performed in 
Ref.~\cite{Bubanja:2023nrd}, 
 corresponds to a nearly flat behavior 
with energy of the intrinsic $k_T$. 
We have already presented a 
discussion of this energy behavior, and  
comparison with the 
collinear parton-shower behavior,  
in our article~\cite{Moureaux:2025cyi}.  
We do not discuss this any further now, 
and refer the reader to 
Ref.~\cite{Moureaux:2025cyi}, and references therein,    
for this topic.

 We regard the results of this paper as providing 
a useful advance in understanding  the 
transverse momentum of vector bosons 
produced in the region of forward rapidities probed by the LHCb experiment. This  region 
is relevant, as mentioned earlier, 
for studies of small-$x$ 
QCD~\cite{Caucal:2025xxh,Hautmann:2012sh,Motyka:2016lta,Taels:2023czt}.    Since the PB TMD approach 
can  be applied to the production of 
jets~\cite{BermudezMartinez:2022bpj,Abdulhamid:2021xtt,Yang:2022qgk}, it is worth  noting  that the analysis 
 presented in this paper could also be helpful 
to address forward jets in hadronic 
collisions~\cite{Deak:2011ga,Deak:2010gk,Deak:2009xt,vanHameren:2023oiq,Deganutti:2023qct,Altinoluk:2023hfz,Wang:2022zdu,Caucal:2025zkl}  
as well as jets in 
small-$x$ lepton-hadron 
scattering~\cite{Hautmann:2007yok,Taels:2022tza,Caucal:2023fsf}.

\subsection{Perturbative matching}
\label{sec:matching}

As discussed in 
Sec.~\ref{sec:methodology-pred}, 
the  study performed in this paper 
uses theoretical predictions for the 
$Z$-boson $p_T$ distributions based 
on the perturbative 
matching~\cite{BermudezMartinez:2019anj,Yang:2022qgk} of next-to-leading-order
matrix elements from 
the MCatNLO  event generator~\cite{Alwall:2014hca} with 
PB TMD~\cite{Hautmann:2017xtx,Hautmann:2017fcj} evolution  
at next-to-leading logarithmic accuracy, implemented in the 
{\sc Cascade}  parton-shower Monte Carlo event 
 generator~\cite{CASCADE:2021bxe}. 
 
The matching is characterised by 
different perturbative scales, which are 
understood to be of the order of the 
hard scales of the process but are otherwise 
arbitrary, and whose variations may be 
used to evaluate theoretical uncertainties 
associated with higher 
fixed orders and/or TMD logarithms. 
The extraction of  non-perturbative 
TMD effects and intrinsic-$k_T$ 
parameters may well depend on details in 
the treatment of the 
perturbative theory; in particular, 
it may depend on the choice of the 
scales for  the perturbative matching. 

To our knowledge, this issue has never 
been addressed in any previous 
study of TMD phenomenology, whether 
based on CSS or PB approaches. 
In this subsection we perform an 
exploratory study of this issue.   
This study is not intended by any means 
to be exhaustive, but rather it is intended 
to draw 
attention to this topic and point to 
directions for future investigations. 

In the PB TMD 
matching~\cite{BermudezMartinez:2019anj,Yang:2022qgk}, 
the matching scale $\mu_m$ is 
set within the MCatNLO generator 
 and is designed to 
consistently avoid double 
counting of parton emissions 
between the hard matrix elements and the 
TMD evolution. 
A common feature often 
observed in the literature on 
precision vector-boson phenomenology, associated 
with procedures of matching of 
fixed-order perturbation theory 
with collinear parton showers 
(as in the original 
MCatNLO method) or with 
TMD evolution (as in the PB TMD method), 
is a bump in the theory/data ratio plot  
in the region of 
vector-boson transverse momenta 
around the matching scale $\mu_m$. 
See, e.g., the theory/data ratio plots in Refs.~\cite{ATLAS:2015iiu,CMS:2022ubq}, in Ref.~\cite{BermudezMartinez:2019anj}, as well as 
Figs.~\ref{fig:MEPBset2vsNNPDF}, \ref{fig:predictions-QEDonOff}  
earlier in this paper. 

\begin{figure}
    \centering
    \includegraphics[width=0.75\linewidth]{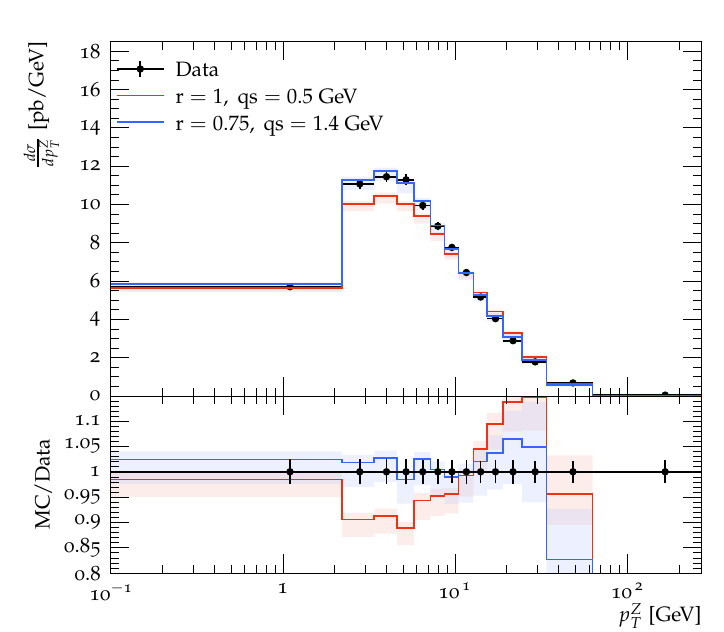}
    \caption{DY prediction obtained by rescaling the matching scale at the level of the Cascade generator.
    }
    \label{fig:ScaleFactorMatchingScale-1}
\end{figure}

Besides $\mu_m$, 
the minimum scale of the hard 
process $\mu_k$  
is passed on to the {\sc Cascade}  Monte Carlo 
generator, and events are rejected 
in which  
at least one of the partons resulting from 
TMD evolution and initiating the 
hard scattering has transverse momentum 
higher than $\mu_k$.  In most applications, 
$ r \equiv \mu_k / \mu_m = 1$,  and 
all previous results in this paper have been 
obtained under this condition. Different 
choices for the scaling parameter 
$r$, however, are possible in 
the  {\sc Cascade}  Monte Carlo,  
influencing the shape 
of the $p_T$ distribution. 
As an example, 
in Fig.~\ref{fig:ScaleFactorMatchingScale-1} 
we illustrate this by comparing the 
result obtained 
with $r=1$ and $q_s = 0.5$ GeV 
 with the result obtained by reducing the 
scaling parameter to $r = 0.75$ and 
increasing the Gaussian width to $q_s = 1.4$ 
GeV. We see that in the latter case the bump 
in the intermediate $p_T$ region near the 
matching scale 
is much reduced and the shape in the 
theory/data ratio plot  is much flatter.

\begin{figure}
    \centering
    \includegraphics[width=0.75\linewidth]{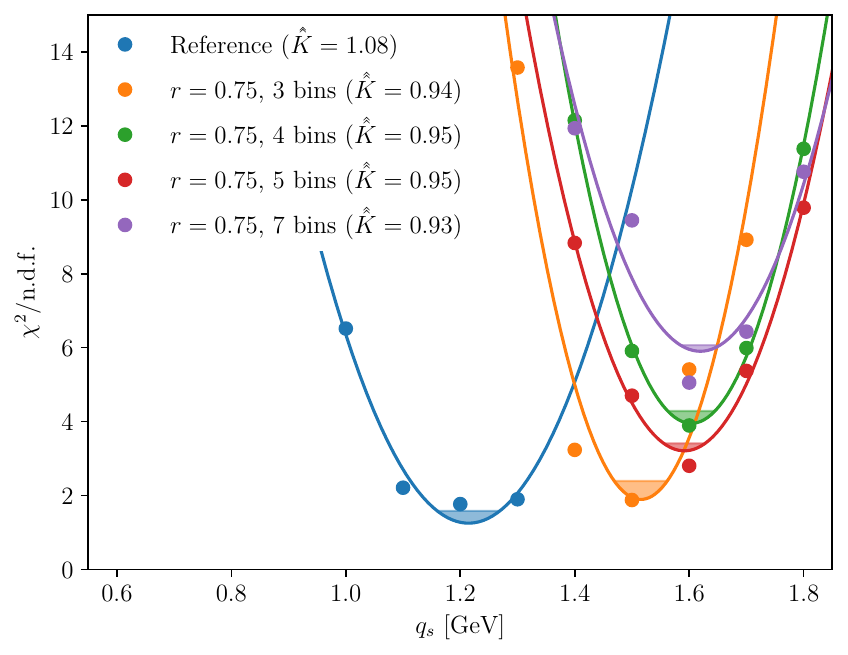}
    \caption{
        $\chi^2/\rm{n.d.f.}$ vs $q_s$ obtained by setting $r=0.75$ at the level of Cascade generator, compared
        to $r=1$ with four bins (``Reference'').
        The photon-induced background is subtracted from all predictions and QED FSR is turned off in both cases.
   }
    \label{fig:ScaleFactorMatchingScale-2}
\end{figure}

Having verified the influence of the 
 matching scale parameter $r$ on 
the $p_T$  shape, 
in Fig.~\ref{fig:ScaleFactorMatchingScale-2} 
we investigate the impact of $r$ on the 
extraction of the intrinsic-$k_T$ parameter 
$q_s$ from fits to experimental data. 
To this end, we repeat the 
analysis done previously 
in Fig.~\ref{fig:KfactorWithCorrelation-QEDonOff},   
by subtracting the photon-induced background 
and turning off the QED final state radiation,  
but now with $r = 0.75$ (instead of $r=1$). 
We note that the change in the best-fit $q_s$ value due to the modified matching scale 
parameter is significant, and that due to the 
flatter $p_T$  shape the result 
is stable with respect to the number of bins, i.e., 
hardly any change is observed when 
including 4, 5 or 7 bins.

As stated earlier in this subsection, by no means do we regard the studies reported 
in Figs.~\ref{fig:ScaleFactorMatchingScale-1} and~\ref{fig:ScaleFactorMatchingScale-2} as exhaustive or conclusive. We however believe that 
they indicate that the matching 
method~\cite{BermudezMartinez:2019anj,Yang:2022qgk} has the capability, not yet fully 
explored, to improve the description of the 
shape of the vector-boson spectra in the intermediate $p_T$ region, around the matching scale $\mu_m$,   and that this indirectly 
influences the  intrinsic-$k_T$ extraction. 
We thus regard further studies of the 
matching as relevant to enhance the 
quality of future TMD determinations. 

 We also note that the perturbative NLO 
matching (and so the possibility to enlarge the 
analysis region toward higher $p_T$)  
is a feature distinguishing the approach to 
the extraction of TMD distributions 
employed in this paper. This may be  compared, on one hand, 
with other TMD extractions in the literature, e.g.~\cite{Bacchetta:2024qre,Moos:2025sal,Barry:2023qqh}, which 
are purely based on low-$p_T$ resummation 
and do not incorporate any matching with 
fixed-order perturbation theory; and, on 
the other hand, with calculations, e.g.~\cite{Camarda:2022qdg,Billis:2024dqq,Camarda:2025lbt}, which do incorporate matching with perturbative fixed orders but do not aim to extract TMD parton distributions. Besides  determinations of TMD distributions, the perturbative matching may  
also be relevant for the short-distance 
(small-$b$) part of the 
Collins-Soper kernel, whose determination 
from experimental DY data can be carried out 
jointly with that of the TMD distributions.

\section{Conclusion}
\label{sec13}
 
In the last few years, 
 precision measurements of the 
transverse-momentum spectra 
of electroweak gauge bosons at the 
LHC and lower-energy experiments 
have enabled one to access 
non-perturbative QCD contributions 
associated with proton's transverse structure. 
In particular, extractions have been carried out 
of the intrinsic transverse 
momentum $k_T$ of partons, 
controlling the TMD 
parton distribution 
for low mass scales in the QCD evolution. 

In this paper we 
have focused on the comparison of 
such extractions from LHCb measurements of 
$Z$-boson spectra in the forward rapidity region 
and from analogous CMS 
measurements in the central rapidity region. 
We have pointed out the importance of taking into account 
correlated uncertainties across 
transverse-momentum bins in these extractions.  
We have illustrated the effect of correlations by 
performing fits to $Z$-boson experimental data 
within a TMD parton-branching analysis. 

In addition, 
we have examined a number of other physical effects which 
potentially influence the determination of 
non-perturbative TMD parameters, including  
final-state electromagnetic radiation, photon-photon 
background,  
collinear-PDF contributions  in the forward region, 
perturbative NLO  matching. 

We have found that in the TMD parton-branching 
determinations  the non-perturbative 
intrinsic-$k_T$  parameters  
extracted from  LHCb forward data 
and CMS central data at 13\,TeV are 
consistent to a level of less than 2$\sigma$, and 
near 1\,GeV.  
This value 
is in striking contrast with 
determinations 
from underlying-event tuning of 
collinear parton-shower Monte 
Carlo event generators to 13\,TeV 
vector-boson data, 
which require intrinsic-$k_T$ values of about 
3\,GeV.

\vskip 0.5 cm 
\noindent {\bf Acknowledgments}. 
We thank  
L.~Han, 
H.~Jung, H.~Li, 
M.~Mangano, N.~Raicevic, P.~Taels, M.~Xu and H.~Yin 
 for fruitful discussions,
as well as B.~Bilin for assistance in generating the photon-induced background.
 LF is supported by the ``Fonds de la Recherche Scientifique - FNRS''.  
AL acknowledges funding by Research Foundation-Flanders 
FWO (application number: 1278325N). 
LM acknowledges support by the DFG
under Germany’s Excellence Strategy 390833306 – EXC 2121: Quantum Universe.

\bibliographystyle{JHEP}
\bibliography{jh-bibl}

@article{Bubanja:2023nrd,
    author = "Bubanja, I. and others",
    title = "{The small $k_{\textrm{T}}$ region in Drell-Yan production at next-to-leading order with the parton branching method}",
    eprint = "2312.08655",
    archivePrefix = "arXiv",
    primaryClass = "hep-ph",
    reportNumber = "DESY-23-209",
    doi = "10.1140/epjc/s10052-024-12507-0",
    journal = "Eur. Phys. J. C",
    volume = "84",
    number = "2",
    pages = "154",
    year = "2024"
}

@article{Hautmann:2017xtx,
    author = "Hautmann, F. and Jung, H. and Lelek, A. and Radescu, V. and Zlebcik, R.",
    title = "{Soft-gluon resolution scale in QCD evolution equations}",
    eprint = "1704.01757",
    archivePrefix = "arXiv",
    primaryClass = "hep-ph",
    reportNumber = "DESY-16-174",
    doi = "10.1016/j.physletb.2017.07.005",
    journal = "Phys. Lett. B",
    volume = "772",
    pages = "446--451",
    year = "2017"
}

@article{Hautmann:2017fcj,
    author = "Hautmann, F. and Jung, H. and Lelek, A. and Radescu, V. and Zlebcik, R.",
    title = "{Collinear and TMD Quark and Gluon Densities from Parton Branching Solution of QCD Evolution Equations}",
    eprint = "1708.03279",
    archivePrefix = "arXiv",
    primaryClass = "hep-ph",
    reportNumber = "DESY-17-118",
    doi = "10.1007/JHEP01(2018)070",
    journal = "JHEP",
    volume = "01",
    pages = "070",
    year = "2018"
}

@article{BermudezMartinez:2019anj,
    author = "Bermudez Martinez, A. and others",
    title = "{Production of Z-bosons in the parton branching method}",
    eprint = "1906.00919",
    archivePrefix = "arXiv",
    primaryClass = "hep-ph",
    reportNumber = "DESY-19-087, CERN-TH-2019-095, DESY 19-087",
    doi = "10.1103/PhysRevD.100.074027",
    journal = "Phys. Rev. D",
    volume = "100",
    number = "7",
    pages = "074027",
    year = "2019"
}

@article{CASCADE:2021bxe,
    author = "Baranov, S. and others",
    collaboration = "CASCADE",
    title = "{CASCADE3 A Monte Carlo event generator based on TMDs}",
    eprint = "2101.10221",
    archivePrefix = "arXiv",
    primaryClass = "hep-ph",
    reportNumber = "DESY-21-005",
    doi = "10.1140/epjc/s10052-021-09203-8",
    journal = "Eur. Phys. J. C",
    volume = "81",
    number = "5",
    pages = "425",
    year = "2021"
}

@article{LHCb:2021huf,
    author = "Aaij, R. and others",
    collaboration = "LHCb",
    title = "{Precision measurement of forward $Z$ boson production in proton-proton collisions at $\sqrt{s} = 13$ TeV}",
    eprint = "2112.07458",
    archivePrefix = "arXiv",
    primaryClass = "hep-ex",
    reportNumber = "LHCb-PAPER-2021-037, CERN-EP-2021-246",
    doi = "10.1007/JHEP07(2022)026",
    journal = "JHEP",
    volume = "07",
    pages = "026",
    year = "2022"
}

@article{CMS:2022ubq,
    author = "Tumasyan, Armen and others",
    collaboration = "CMS",
    title = "{Measurement of the mass dependence of the transverse momentum of lepton pairs in Drell-Yan production in proton-proton collisions at $\sqrt{s}$ = 13 TeV}",
    eprint = "2205.04897",
    archivePrefix = "arXiv",
    primaryClass = "hep-ex",
    reportNumber = "CMS-SMP-20-003, CERN-EP-2022-053",
    doi = "10.1140/epjc/s10052-023-11631-7",
    journal = "Eur. Phys. J. C",
    volume = "83",
    number = "7",
    pages = "628",
    year = "2023"
}

@article{NNPDF:2014otw,
    author = "Ball, Richard D. and others",
    collaboration = "NNPDF",
    title = "{Parton distributions for the LHC Run II}",
    eprint = "1410.8849",
    archivePrefix = "arXiv",
    primaryClass = "hep-ph",
    reportNumber = "EDINBURGH-2014-15, IFUM-1034-FT, CERN-PH-TH-2013-253, OUTP-14-11P, CAVENDISH-HEP-14-11",
    doi = "10.1007/JHEP04(2015)040",
    journal = "JHEP",
    volume = "04",
    pages = "040",
    year = "2015"
}

@article{CASCADE:2010clj,
    author = "Jung, H. and others",
    collaboration = "CASCADE",
    title = "{The CCFM Monte Carlo generator CASCADE version 2.2.03}",
    eprint = "1008.0152",
    archivePrefix = "arXiv",
    primaryClass = "hep-ph",
    reportNumber = "DESY-10-107",
    doi = "10.1140/epjc/s10052-010-1507-z",
    journal = "Eur. Phys. J. C",
    volume = "70",
    pages = "1237--1249",
    year = "2010"
}

@article{Corcella:2002jc,
    author = "Corcella, G. and Knowles, I. G. and Marchesini, G. and Moretti, S. and Odagiri, K. and Richardson, P. and Seymour, M. H. and Webber, B. R.",
    title = "{HERWIG 6.5 release note}",
    eprint = "hep-ph/0210213",
    archivePrefix = "arXiv",
    reportNumber = "CAVENDISH-HEP-02-17, DAMTP-2002-124, KEK-TH-850, MPI-PHT-2002-55, CERN-TH-2002-270, IPPP-02-58, MC-TH-2002-7",
    month = "10",
    year = "2002"
}

@article{Corcella:2000bw,
    author = "Corcella, G. and Knowles, I. G. and Marchesini, G. and Moretti, S. and Odagiri, K. and Richardson, P. and Seymour, M. H. and Webber, B. R.",
    title = "{HERWIG 6: An Event generator for hadron emission reactions with interfering gluons (including supersymmetric processes)}",
    eprint = "hep-ph/0011363",
    archivePrefix = "arXiv",
    reportNumber = "CAVENDISH-HEP-99-03, CERN-TH-2000-284, RAL-TR-2000-048",
    doi = "10.1088/1126-6708/2001/01/010",
    journal = "JHEP",
    volume = "01",
    pages = "010",
    year = "2001"
}

@article{Sjostrand:2006za,
    author = "Sjostrand, Torbjorn and Mrenna, Stephen and Skands, Peter Z.",
    title = "{PYTHIA 6.4 Physics and Manual}",
    eprint = "hep-ph/0603175",
    archivePrefix = "arXiv",
    reportNumber = "FERMILAB-PUB-06-052-CD-T, LU-TP-06-13",
    doi = "10.1088/1126-6708/2006/05/026",
    journal = "JHEP",
    volume = "05",
    pages = "026",
    year = "2006"
}

@article{Martinez:2024twn,
    author = "Bermudez Martinez, A.  and others",
    title = "{The Parton Branching Sudakov and its relation to CSS}",
    doi = "10.22323/1.449.0270",
    journal = "PoS",
    volume = "EPS-HEP2023",
    pages = "270",
    year = "2024"
}

@article{Martinez:2024mou,
    author = "Bermudez Martinez, A. and others",
    title = "{Soft-gluon coupling and the TMD parton branching Sudakov form factor}",
    eprint = "2412.21116",
    archivePrefix = "arXiv",
    primaryClass = "hep-ph",
    doi = "10.1016/j.physletb.2025.139762",
    journal = "Phys. Lett. B",
    volume = "868",
    pages = "139762",
    year = "2025"
}

@Article{Banfi:2018mcq,
  author        = {Banfi, Andrea and El-Menoufi, Basem Kamal and Monni, Pier Francesco},
  title         = {{The Sudakov radiator for jet observables and the soft physical coupling}},
  journal       = {JHEP},
  year          = {2019},
  volume        = {01},
  pages         = {083},
  archiveprefix = {arXiv},
  doi           = {10.1007/JHEP01(2019)083},
  eprint        = {1807.11487},
  primaryclass  = {hep-ph},
}

@Article{Catani:2019rvy,
  author        = {Catani, Stefano and De Florian, Daniel and Grazzini, Massimiliano},
  title         = {{Soft-gluon effective coupling and cusp anomalous dimension}},
  journal       = {Eur. Phys. J. C},
  year          = {2019},
  volume        = {79},
  number        = {8},
  pages         = {685},
  archiveprefix = {arXiv},
  doi           = {10.1140/epjc/s10052-019-7174-9},
  eprint        = {1904.10365},
  primaryclass  = {hep-ph},
}

@Article{BermudezMartinez:2020tys,
  author        = {Bermudez Martinez, A. and others},
  title         = {{The transverse momentum spectrum of low mass Drell-Yan production at next-to-leading order in the parton branching method}},
  journal       = {Eur. Phys. J. C},
  year          = {2020},
  volume        = {80},
  number        = {7},
  pages         = {598},
  archiveprefix = {arXiv},
  doi           = {10.1140/epjc/s10052-020-8136-y},
  eprint        = {2001.06488},
  primaryclass  = {hep-ph},
  reportnumber  = {DESY 20-006, DESY-20-006},
}

@Article{BermudezMartinez:2018fsv,
  author        = {Bermudez Martinez, A. and others},
  title         = {{Collinear and TMD parton densities from fits to precision DIS measurements in the parton branching method}},
  journal       = {Phys. Rev. D},
  year          = {2019},
  volume        = {99},
  number        = {7},
  pages         = {074008},
  archiveprefix = {arXiv},
  doi           = {10.1103/PhysRevD.99.074008},
  eprint        = {1804.11152},
  primaryclass  = {hep-ph},
  reportnumber  = {DESY 18-042, DESY-18-042},
}

@Article{Yang:2022qgk,
  author        = {Yang, H. and others},
  title         = {{Back-to-back azimuthal correlations in $\mathrm {Z} +$jet events at high transverse momentum in the TMD parton branching method at next-to-leading order}},
  journal       = {Eur. Phys. J. C},
  year          = {2022},
  volume        = {82},
  number        = {8},
  pages         = {755},
  archiveprefix = {arXiv},
  doi           = {10.1140/epjc/s10052-022-10715-0},
  eprint        = {2204.01528},
  primaryclass  = {hep-ph},
  reportnumber  = {DESY-22-025, CERN-TH-2022-113},
}

@article{Abdulhamid:2021xtt,
    author = "Abdulhamid, M. I. and others",
    title = "{Azimuthal correlations of high transverse momentum jets at next-to-leading order in the parton branching method}",
    eprint = "2112.10465",
    archivePrefix = "arXiv",
    primaryClass = "hep-ph",
    reportNumber = "DESY-21-219, LU-TP 21-53",
    doi = "10.1140/epjc/s10052-022-09997-1",
    journal = "Eur. Phys. J. C",
    volume = "82",
    number = "1",
    pages = "36",
    year = "2022"
}

@Article{Buckley:2010ar,
  author        = {Buckley, Andy and Butterworth, Jonathan and Grellscheid, David and Hoeth, Hendrik and Lonnblad, Leif and Monk, James and Schulz, Holger and Siegert, Frank},
  title         = {{Rivet user manual}},
  journal       = {Comput. Phys. Commun.},
  year          = {2013},
  volume        = {184},
  pages         = {2803--2819},
  archiveprefix = {arXiv},
  doi           = {10.1016/j.cpc.2013.05.021},
  eprint        = {1003.0694},
  primaryclass  = {hep-ph},
  reportnumber  = {MCNET-10-03, MCnet/10/03},
}

@Article{Alwall:2014hca,
  author        = {Alwall, J. and Frederix, R. and Frixione, S. and Hirschi, V. and Maltoni, F. and Mattelaer, O. and Shao, H. -S. and Stelzer, T. and Torrielli, P. and Zaro, M.},
  title         = {{The automated computation of tree-level and next-to-leading order differential cross sections, and their matching to parton shower simulations}},
  journal       = {JHEP},
  year          = {2014},
  volume        = {07},
  pages         = {079},
  archiveprefix = {arXiv},
  doi           = {10.1007/JHEP07(2014)079},
  eprint        = {1405.0301},
  primaryclass  = {hep-ph},
  reportnumber  = {CERN-PH-TH-2014-064, CP3-14-18, LPN14-066, MCNET-14-09, ZU-TH-14-14},
}

@article{Angeles-Martinez:2015sea,
    author = "Angeles-Martinez, R. and others",
    title = "{Transverse Momentum Dependent (TMD) parton distribution functions: status and prospects}",
    eprint = "1507.05267",
    archivePrefix = "arXiv",
    primaryClass = "hep-ph",
    reportNumber = "DESY-15-111, NIKHEF-2015-023, RAL-P-2015-006, JLAB-THY-15-2020",
    doi = "10.5506/APhysPolB.46.2501",
    journal = "Acta Phys. Polon. B",
    volume = "46",
    number = "12",
    pages = "2501--2534",
    year = "2015"
}

@article{Abdulov:2021ivr,
    author = "Abdulov, N. A. and others",
    title = "{TMDlib2 and TMDplotter: a platform for 3D hadron structure studies}",
    eprint = "2103.09741",
    archivePrefix = "arXiv",
    primaryClass = "hep-ph",
    reportNumber = "DESY 21-026, DESY-21-026, IFJPAN-IV-2021-4, JLAB-THY-21-3337",
    doi = "10.1140/epjc/s10052-021-09508-8",
    journal = "Eur. Phys. J. C",
    volume = "81",
    number = "8",
    pages = "752",
    year = "2021"
}

@article{BermudezMartinez:2021lxz,
    author = "Bermudez Martinez, A. and Hautmann, F. and Mangano, M.",
    title = "{TMD evolution and multi-jet merging}",
    eprint = "2107.01224",
    archivePrefix = "arXiv",
    primaryClass = "hep-ph",
    doi = "10.1016/j.physletb.2021.136700",
    journal = "Phys. Lett. B",
    volume = "822",
    pages = "136700",
    year = "2021"
}

@article{BermudezMartinez:2022bpj,
    author = "Bermudez Martinez, A. and Hautmann, F. and Mangano, M.",
    title = "{Multi-jet merging with TMD parton branching}",
    eprint = "2208.02276",
    archivePrefix = "arXiv",
    primaryClass = "hep-ph",
    reportNumber = "CERN-TH-2022-131, DESY-22-133",
    doi = "10.1007/JHEP09(2022)060",
    journal = "JHEP",
    volume = "09",
    pages = "060",
    year = "2022"
}

@inproceedings{BermudezMartinez:2021zlg,
    author = "Bermudez Martinez, A. and Hautmann, F. and Mangano, M.",
    title = "{Multi-jet physics at high-energy colliders and TMD parton evolution}",
    eprint = "2109.08173",
    archivePrefix = "arXiv",
    primaryClass = "hep-ph",
    month = "9",
    year = "2021"
}

@article{Zhan:2024lym,
    author = "Zhan, Wenxiao and others",
    title = "{Coarse-grained binning in Drell-Yan transverse momentum spectra}",
    eprint = "2412.19060",
    archivePrefix = "arXiv",
    primaryClass = "hep-ph",
    doi = "10.1103/PhysRevD.111.036018",
    journal = "Phys. Rev. D",
    volume = "111",
    number = "3",
    pages = "036018",
    year = "2025"
}

@article{Caucal:2025mth,
    author = "Caucal, Paul and Iancu, Edmond and Salazar, Farid and Yuan, Feng",
    title = "{Gluon splitting at small x: a unified derivation for the JIMWLK, DGLAP and CSS equations}",
    eprint = "2510.08454",
    archivePrefix = "arXiv",
    primaryClass = "hep-ph",
    doi = "10.1007/JHEP03(2026)198",
    journal = "JHEP",
    volume = "03",
    pages = "198",
    year = "2026"
}

@article{Caucal:2025xxh,
    author = "Caucal, Paul and Morales, Marcos Guerrero and Iancu, Edmond and Salazar, Farid and Yuan, Feng",
    title = "{Unveiling the sea: universality of the transverse momentum dependent quark distributions at small x}",
    eprint = "2503.16162",
    archivePrefix = "arXiv",
    primaryClass = "hep-ph",
    doi = "10.1016/j.physletb.2026.140271",
    journal = "Phys. Lett. B",
    volume = "874",
    pages = "140271",
    year = "2026"
}

@article{Caucal:2024bae,
    author = "Caucal, Paul and Iancu, Edmond",
    title = "{Evolution of the transverse-momentum dependent gluon distribution at small x}",
    eprint = "2406.04238",
    archivePrefix = "arXiv",
    primaryClass = "hep-ph",
    doi = "10.1103/PhysRevD.111.074008",
    journal = "Phys. Rev. D",
    volume = "111",
    number = "7",
    pages = "074008",
    year = "2025"
}

@article{Duan:2024qev,
    author = "Duan, Haowu and Kovner, Alex and Lublinsky, Michael",
    title = "{Born-Oppenheimer renormalization group for high energy scattering: CSS, DGLAP and all that}",
    eprint = "2412.05097",
    archivePrefix = "arXiv",
    primaryClass = "hep-ph",
    doi = "10.1007/JHEP08(2025)137",
    journal = "JHEP",
    volume = "08",
    pages = "137",
    year = "2025"
}

@article{Duan:2024qck,
    author = "Duan, Haowu and Kovner, Alex and Lublinsky, Michael",
    title = "{Born-Oppenheimer renormalization group for high energy scattering: the setup and the wave function}",
    eprint = "2412.05085",
    archivePrefix = "arXiv",
    primaryClass = "hep-ph",
    doi = "10.1007/JHEP08(2025)136",
    journal = "JHEP",
    volume = "08",
    pages = "136",
    year = "2025"
}

@article{Mukherjee:2023snp,
    author = "Mukherjee, Swagato and Skokov, Vladimir V. and Tarasov, Andrey and Tiwari, Shaswat",
    title = "{Unified description of DGLAP, CSS, and BFKL evolution: TMD factorization bridging large and small x}",
    eprint = "2311.16402",
    archivePrefix = "arXiv",
    primaryClass = "hep-ph",
    doi = "10.1103/PhysRevD.109.034035",
    journal = "Phys. Rev. D",
    volume = "109",
    number = "3",
    pages = "034035",
    year = "2024"
}

@article{Altinoluk:2025ewj,
    author = "Altinoluk, Tolga and Beuf, Guillaume and Jalilian-Marian, Jamal",
    title = "{One-loop renormalization of quark TMD in the light-cone gauge: CSS evolution}",
    eprint = "2505.20467",
    archivePrefix = "arXiv",
    primaryClass = "hep-ph",
    month = "5",
    year = "2025"
}

@article{Taels:2023czt,
    author = "Taels, Pieter",
    title = "{Forward production of a Drell-Yan pair and a jet at small x at next-to-leading order}",
    eprint = "2308.02449",
    archivePrefix = "arXiv",
    primaryClass = "hep-ph",
    doi = "10.1007/JHEP01(2024)005",
    journal = "JHEP",
    volume = "01",
    pages = "005",
    year = "2024"
}

@article{Moos:2025sal,
    author = "Moos, Valentin and Scimemi, Ignazio and Vladimirov, Alexey and Zurita, Pia",
    title = "{Determination of unpolarized TMD distributions from the fit of Drell-Yan and SIDIS data at N$^{4}$LL}",
    eprint = "2503.11201",
    archivePrefix = "arXiv",
    primaryClass = "hep-ph",
    reportNumber = "IPARCOS-UCM-25-018",
    doi = "10.1007/JHEP11(2025)134",
    journal = "JHEP",
    volume = "11",
    pages = "134",
    year = "2025"
}

@article{Bacchetta:2024qre,
    author = "Bacchetta, Alessandro and Bertone, Valerio and Bissolotti, Chiara and Bozzi, Giuseppe and Cerutti, Matteo and Delcarro, Filippo and Radici, Marco and Rossi, Lorenzo and Signori, Andrea",
    collaboration = "MAP (Multi-dimensional Analyses of Partonic distributions)",
    title = "{Flavor dependence of unpolarized quark transverse momentum distributions from a global fit}",
    eprint = "2405.13833",
    archivePrefix = "arXiv",
    primaryClass = "hep-ph",
    reportNumber = "JLAB-THY-24-4066",
    doi = "10.1007/JHEP08(2024)232",
    journal = "JHEP",
    volume = "08",
    pages = "232",
    year = "2024"
}

@article{CMS:2024goo,
    author = "Hayrapetyan, Aram and others",
    collaboration = "CMS",
    title = "{Energy-scaling behavior of intrinsic transverse-momentum parameters in Drell-Yan simulation}",
    eprint = "2409.17770",
    archivePrefix = "arXiv",
    primaryClass = "hep-ph",
    reportNumber = "CMS-GEN-22-001, CERN-EP-2024-216",
    doi = "10.1103/PhysRevD.111.072003",
    journal = "Phys. Rev. D",
    volume = "111",
    number = "7",
    pages = "072003",
    year = "2025"
}

@article{Moureaux:2025cyi,
    author = "Moureaux, Louis and others",
    title = "{Intrinsic kT and soft gluons in Monte Carlo event generators}",
    eprint = "2511.23291",
    archivePrefix = "arXiv",
    primaryClass = "hep-ph",
    doi = "10.22323/1.485.0256",
    journal = "PoS",
    volume = "EPS-HEP2025",
    pages = "256",
    year = "2026"
}

@article{Hautmann:2025fkw,
    author = "Hautmann, F. and others",
    title = "{Collinear and TMD distributions with dynamical soft-gluon resolution scale}",
    eprint = "2502.19380",
    archivePrefix = "arXiv",
    primaryClass = "hep-ph",
    reportNumber = "DESY-25-031",
    doi = "10.1007/JHEP06(2025)192",
    journal = "JHEP",
    volume = "06",
    pages = "192",
    year = "2025"
}

@article{Bubanja:2024puv,
    author = "Bubanja, I. and others",
    title = "{Center-of-mass energy dependence of intrinsic-$k_{\textrm{T}}$ distributions obtained from Drell{\textendash}Yan production}",
    eprint = "2404.04088",
    archivePrefix = "arXiv",
    primaryClass = "hep-ph",
    reportNumber = "DESY-24-049",
    doi = "10.1140/epjc/s10052-025-14021-3",
    journal = "Eur. Phys. J. C",
    volume = "85",
    number = "3",
    pages = "278",
    year = "2025"
}

@article{Bellm:2015jjp,
    author = "Bellm, Johannes and others",
    title = "{Herwig 7.0/Herwig++ 3.0 release note}",
    eprint = "1512.01178",
    archivePrefix = "arXiv",
    primaryClass = "hep-ph",
    reportNumber = "CERN-PH-TH-2015-289, MAN-HEP-2015-15, IFJPAN-IV-2015-13, KA-TP-18-2015, DCPT-15-142, MCNET-15-28, IPPP-15-71, HERWIG-2015-01",
    doi = "10.1140/epjc/s10052-016-4018-8",
    journal = "Eur. Phys. J. C",
    volume = "76",
    number = "4",
    pages = "196",
    year = "2016"
}

@article{Sjostrand:2014zea,
    author = {Sj\"ostrand, Torbj\"orn and Ask, Stefan and Christiansen, Jesper R. and Corke, Richard and Desai, Nishita and Ilten, Philip and Mrenna, Stephen and Prestel, Stefan and Rasmussen, Christine O. and Skands, Peter Z.},
    title = "{An introduction to PYTHIA 8.2}",
    eprint = "1410.3012",
    archivePrefix = "arXiv",
    primaryClass = "hep-ph",
    reportNumber = "LU-TP-14-36, MCNET-14-22, CERN-PH-TH-2014-190, FERMILAB-PUB-14-316-CD, DESY-14-178, SLAC-PUB-16122",
    doi = "10.1016/j.cpc.2015.01.024",
    journal = "Comput. Phys. Commun.",
    volume = "191",
    pages = "159--177",
    year = "2015"
}

@article{Azzi:2019yne,
    author = "Azzi, P. and others",
    editor = "Dainese, Andrea and Mangano, Michelangelo and Meyer, Andreas B. and Nisati, Aleandro and Salam, Gavin and Vesterinen, Mika Anton",
    title = "{Report from Working Group 1}: {Standard Model Physics at the HL-LHC and HE-LHC}",
    eprint = "1902.04070",
    archivePrefix = "arXiv",
    primaryClass = "hep-ph",
    reportNumber = "CERN-LPCC-2018-03",
    doi = "10.23731/CYRM-2019-007.1",
    journal = "CERN Yellow Rep. Monogr.",
    volume = "7",
    pages = "1--220",
    year = "2019"
}

@article{LHeC:2020van,
    author = "Agostini, P. and others",
    collaboration = "LHeC, FCC-he Study Group",
    title = "{The Large Hadron-Electron Collider at the HL-LHC}",
    eprint = "2007.14491",
    archivePrefix = "arXiv",
    primaryClass = "hep-ex",
    reportNumber = "CERN-ACC-Note-2020-0002, JLAB-ACP-20-3180",
    doi = "10.1088/1361-6471/abf3ba",
    journal = "J. Phys. G",
    volume = "48",
    number = "11",
    pages = "110501",
    year = "2021"
}

@article{FCC:2018byv,
    author = "Abada, A. and others",
    collaboration = "FCC",
    title = "{FCC Physics Opportunities}: {Future Circular Collider Conceptual Design Report Volume 1}",
    reportNumber = "CERN-ACC-2018-0056",
    doi = "10.1140/epjc/s10052-019-6904-3",
    journal = "Eur. Phys. J. C",
    volume = "79",
    number = "6",
    pages = "474",
    year = "2019"
}

@article{Proceedings:2020eah,
    author = "Hatta, Yoshitaka and others",
    title = "{Proceedings, Probing Nucleons and Nuclei in High Energy Collisions: Dedicated to the Physics of the Electron Ion Collider}: {Seattle (WA), United States, October 1 - November 16, 2018}",
    eprint = "2002.12333",
    archivePrefix = "arXiv",
    primaryClass = "hep-ph",
    doi = "10.1142/11684",
    publisher = "WSP",
    month = "2",
    year = "2020"
}

@inproceedings{CEPCPhysicsStudyGroup:2022uwl,
    author = "Cheng, Huajie and others",
    collaboration = "CEPC Physics Study Group",
    title = "{The Physics potential of the CEPC. Prepared for the US Snowmass Community Planning Exercise (Snowmass 2021)}",
    booktitle = "{Snowmass 2021}",
    eprint = "2205.08553",
    archivePrefix = "arXiv",
    primaryClass = "hep-ph",
    month = "5",
    year = "2022"
}

@article{Cipriano:2013ooa,
    author = "Cipriano, P. and others",
    title = "{Higgs boson as a gluon trigger}",
    eprint = "1308.1655",
    archivePrefix = "arXiv",
    primaryClass = "hep-ph",
    reportNumber = "DESY-13-139",
    doi = "10.1103/PhysRevD.88.097501",
    journal = "Phys. Rev. D",
    volume = "88",
    number = "9",
    pages = "097501",
    year = "2013"
}

@article{Boer:2024ylx,
    author = {Boer, Dani{\"e}l and others},
    title = "{Physics case for quarkonium studies at the Electron Ion Collider}",
    eprint = "2409.03691",
    archivePrefix = "arXiv",
    primaryClass = "hep-ph",
    doi = "10.1016/j.ppnp.2025.104162",
    journal = "Prog. Part. Nucl. Phys.",
    volume = "142",
    pages = "104162",
    year = "2025"
}

@article{CMS:2024lrd,
    author = "Chekhovsky, Vladimir and others",
    collaboration = "CMS",
    title = "{High-precision measurement of the W boson mass with the CMS experiment at the LHC}",
    eprint = "2412.13872",
    archivePrefix = "arXiv",
    primaryClass = "hep-ex",
    reportNumber = "CMS-SMP-23-002, CERN-EP-2024-308",
    month = "12",
    year = "2024"
}

@article{ATLAS:2023lhg,
    author = "Aad, Georges and others",
    collaboration = "ATLAS",
    title = "{A precise determination of the strong-coupling constant from the recoil of $Z$ bosons with the ATLAS experiment at $\sqrt{s} = 8$ TeV}",
    eprint = "2309.12986",
    archivePrefix = "arXiv",
    primaryClass = "hep-ex",
    month = "9",
    year = "2023"
}

@article{Camarda:2022qdg,
    author = "Camarda, Stefano and Ferrera, Giancarlo and Schott, Matthias",
    title = "{Determination of the strong-coupling constant from the Z-boson transverse-momentum distribution}",
    eprint = "2203.05394",
    archivePrefix = "arXiv",
    primaryClass = "hep-ph",
    doi = "10.1140/epjc/s10052-023-12373-2",
    journal = "Eur. Phys. J. C",
    volume = "84",
    number = "1",
    pages = "39",
    year = "2024"
}

@article{Billis:2024dqq,
    author = "Billis, Georgios and Michel, Johannes K. L. and Tackmann, Frank J.",
    title = "{Drell-Yan transverse-momentum spectra at N$^{3}$LL$^{\prime}$ and approximate N$^{4}$LL with SCETlib}",
    eprint = "2411.16004",
    archivePrefix = "arXiv",
    primaryClass = "hep-ph",
    reportNumber = "DESY-23-081, MIT-CTP 5572, Nikhef 2024-007",
    doi = "10.1007/JHEP02(2025)170",
    journal = "JHEP",
    volume = "02",
    pages = "170",
    year = "2025"
}

@article{Barry:2023qqh,
    author = "Barry, P. C. and Gamberg, L. and Melnitchouk, W. and Moffat, E. and Pitonyak, D. and Prokudin, A. and Sato, N.",
    collaboration = "Jefferson Lab Angular Momentum (JAM)",
    title = "{Tomography of pions and protons via transverse momentum dependent distributions}",
    eprint = "2302.01192",
    archivePrefix = "arXiv",
    primaryClass = "hep-ph",
    reportNumber = "JLAB-THY-23-3749, ADP-23-03/T1212",
    doi = "10.1103/PhysRevD.108.L091504",
    journal = "Phys. Rev. D",
    volume = "108",
    number = "9",
    pages = "L091504",
    year = "2023"
}

@article{Barry:2025glq,
    author = "Barry, P. C. and others",
    title = "{First simultaneous analysis of transverse momentum dependent and collinear parton distributions in the proton}",
    eprint = "2510.13771",
    archivePrefix = "arXiv",
    primaryClass = "hep-ph",
    reportNumber = "JLAB-THY-25-4573, IPARCOS-UCM-25-051",
    month = "10",
    year = "2025"
}

@article{Zaccheddu:2026pyy,
    author = "Zaccheddu, Marco and Gamberg, Leonard and Melnitchouk, Wally and Pitonyak, Daniel and Prokudin, Alexei and Qiu, Jian-Wei and Sato, Nobuo",
    title = "{TMDs in the Lens of Generative AI: A Pixel-Based Approach to Partonic Imaging}",
    eprint = "2605.06606",
    archivePrefix = "arXiv",
    primaryClass = "hep-ph",
    reportNumber = "JLAB-THY-26-4691",
    month = "5",
    year = "2026"
}

@article{Aslan:2024nqg,
    author = "Aslan, F. and Boglione, M. and Gonzalez-Hernandez, J. O. and Rainaldi, T. and Rogers, T. C. and Simonelli, A.",
    title = "{Phenomenology of TMD parton distributions in Drell-Yan and Z0 boson production in a hadron structure oriented approach}",
    eprint = "2401.14266",
    archivePrefix = "arXiv",
    primaryClass = "hep-ph",
    reportNumber = "JLAB-THY-24-3987",
    doi = "10.1103/PhysRevD.110.074016",
    journal = "Phys. Rev. D",
    volume = "110",
    number = "7",
    pages = "074016",
    year = "2024"
}

@article{Cerutti:2026apy,
    author = "Cerutti, Matteo and Simonelli, Andrea",
    title = "{The impact of prescriptions in phenomenological extractions of Transverse Momentum Dependent distributions}",
    eprint = "2603.19088",
    archivePrefix = "arXiv",
    primaryClass = "hep-ph",
    month = "3",
    year = "2026"
}

@article{Camarda:2025lbt,
    author = "Camarda, Stefano and Ferrera, Giancarlo and Rossi, Lorenzo",
    title = "{Drell-Yan lepton pair production at low invariant masses: transverse-momentum resummation and non-perturbative effects in QCD}",
    eprint = "2508.06201",
    archivePrefix = "arXiv",
    primaryClass = "hep-ph",
    doi = "10.1007/JHEP01(2026)150",
    journal = "JHEP",
    volume = "01",
    pages = "150",
    year = "2026"
}

@article{Bollweg:2025iol,
    author = "Bollweg, Dennis and Gao, Xiang and He, Jinchen and Mukherjee, Swagato and Zhao, Yong",
    title = "{Transverse-momentum-dependent pion structures from lattice QCD: Collins-Soper kernel, soft factor, TMDWF, and TMDPDF}",
    eprint = "2504.04625",
    archivePrefix = "arXiv",
    primaryClass = "hep-lat",
    doi = "10.1103/j3n6-8kxy",
    journal = "Phys. Rev. D",
    volume = "112",
    number = "3",
    pages = "034501",
    year = "2025"
}

@article{Bury:2022czx,
    author = "Bury, Marcin and Hautmann, Francesco and Leal-Gomez, Sergio and Scimemi, Ignazio and Vladimirov, Alexey and Zurita, Pia",
    title = "{PDF bias and flavor dependence in TMD distributions}",
    eprint = "2201.07114",
    archivePrefix = "arXiv",
    primaryClass = "hep-ph",
    reportNumber = "UWThPh 2021-29, CERN-TH-2022-126",
    doi = "10.1007/JHEP10(2022)118",
    journal = "JHEP",
    volume = "10",
    pages = "118",
    year = "2022"
}

@article{Hautmann:2021ovt,
    author = "Hautmann, Francesco and Scimemi, Ignazio and Vladimirov, Alexey",
    title = "{Determination of the rapidity evolution kernel from Drell-Yan data at low transverse momenta}",
    eprint = "2109.12051",
    archivePrefix = "arXiv",
    primaryClass = "hep-ph",
    doi = "10.21468/SciPostPhysProc.8.123",
    journal = "SciPost Phys. Proc.",
    volume = "8",
    pages = "123",
    year = "2022"
}

@article{Hautmann:2020cyp,
    author = "Hautmann, Francesco and Scimemi, Ignazio and Vladimirov, Alexey",
    title = "{Non-perturbative contributions to vector-boson transverse momentum spectra in hadronic collisions}",
    eprint = "2002.12810",
    archivePrefix = "arXiv",
    primaryClass = "hep-ph",
    doi = "10.1016/j.physletb.2020.135478",
    journal = "Phys. Lett. B",
    volume = "806",
    pages = "135478",
    year = "2020"
}

@article{Aybat:2011zv,
    author = "Aybat, S. Mert and Rogers, Ted C.",
    title = "{TMD Parton Distribution and Fragmentation Functions with QCD Evolution}",
    eprint = "1101.5057",
    archivePrefix = "arXiv",
    primaryClass = "hep-ph",
    reportNumber = "NIKHEF-2011-001",
    doi = "10.1103/PhysRevD.83.114042",
    journal = "Phys. Rev. D",
    volume = "83",
    pages = "114042",
    year = "2011"
}

@article{Collins:1984kg,
    author = "Collins, John C. and Soper, Davison E. and Sterman, George F.",
    title = "{Transverse Momentum Distribution in Drell-Yan Pair and W and Z Boson Production}",
    reportNumber = "CERN-TH-3923",
    doi = "10.1016/0550-3213(85)90479-1",
    journal = "Nucl. Phys. B",
    volume = "250",
    pages = "199--224",
    year = "1985"
}

@article{Hautmann:2014kza,
    author = {Hautmann, F. and Jung, H. and Kr{\"a}mer, M. and Mulders, P. J. and Nocera, E. R. and Rogers, T. C. and Signori, A.},
    title = "{TMDlib and TMDplotter: library and plotting tools for transverse-momentum-dependent parton distributions}",
    eprint = "1408.3015",
    archivePrefix = "arXiv",
    primaryClass = "hep-ph",
    reportNumber = "DESY-14-059, NIKHEF-2014-024, YITP-SB-14-24",
    doi = "10.1140/epjc/s10052-014-3220-9",
    journal = "Eur. Phys. J. C",
    volume = "74",
    pages = "3220",
    year = "2014"
}

@article{Hautmann:2012sh,
    author = "Hautmann, F. and Hentschinski, M. and Jung, H.",
    title = "{Forward Z-boson production and the unintegrated sea quark density}",
    eprint = "1205.1759",
    archivePrefix = "arXiv",
    primaryClass = "hep-ph",
    reportNumber = "IFT-UAM-CSIC-12-39, LPN12-050, OUTP-12-08-P",
    doi = "10.1016/j.nuclphysb.2012.07.023",
    journal = "Nucl. Phys. B",
    volume = "865",
    pages = "54--66",
    year = "2012"
}

@Article{Hautmann:2022xuc,
  author        = {Hautmann, F. and Hentschinski, M. and Keersmaekers, L. and Kusina, A. and Kutak, K. and Lelek, A.},
  title         = {{A parton branching with transverse momentum dependent splitting functions}},
  journal       = {Phys. Lett. B},
  year          = {2022},
  volume        = {833},
  pages         = {137276},
  archiveprefix = {arXiv},
  doi           = {10.1016/j.physletb.2022.137276},
  eprint        = {2205.15873},
  primaryclass  = {hep-ph},
  reportnumber  = {CERN-TH-2022-087, IFJPAN-IV-2022-8},
}

@article{Collins:1981va,
      author         = "Collins, John C. and Soper, Davison E.",
      title          = "{Back-To-Back Jets: Fourier Transform from B to
                        K-Transverse}",
      journal        = "Nucl. Phys.",
      volume         = "B197",
      year           = "1982",
      pages          = "446-476",
      doi            = "10.1016/0550-3213(82)90453-9",
      reportNumber   = "OITS-153",
      SLACcitation   = "%%CITATION = NUPHA,B197,446;%%"
}

@article{Collins:1981uk,
      author         = "Collins, John C. and Soper, Davison E.",
      title          = "{Back-To-Back Jets in QCD}",
      journal        = "Nucl. Phys.",
      volume         = "B193",
      year           = "1981",
      pages          = "381",
      doi            = "10.1016/0550-3213(81)90339-4",
      note           = "[Erratum: Nucl. Phys.B213,545(1983)]",
      reportNumber   = "OITS-155",
      SLACcitation   = "%%CITATION = NUPHA,B193,381;%%"
}

@article{Avkhadiev:2024mgd,
    author = "Avkhadiev, Artur and Shanahan, Phiala E. and Wagman, Michael L. and Zhao, Yong",
    title = "{Determination of the Collins-Soper Kernel from Lattice QCD}",
    eprint = "2402.06725",
    archivePrefix = "arXiv",
    primaryClass = "hep-lat",
    reportNumber = "FERMILAB-PUB-24-0037-T, MIT-CTP/5677",
    doi = "10.1103/PhysRevLett.132.231901",
    journal = "Phys. Rev. Lett.",
    volume = "132",
    number = "23",
    pages = "231901",
    year = "2024"
}

@Article{LPC:2022ibr,
  author        = {Chu, Min-Huan and others},
  title         = {{Nonperturbative determination of the Collins-Soper kernel from quasitransverse-momentum-dependent wave functions}},
  journal       = {Phys. Rev. D},
  year          = {2022},
  volume        = {106},
  number        = {3},
  pages         = {034509},
  archiveprefix = {arXiv},
  collaboration = {LPC},
  doi           = {10.1103/PhysRevD.106.034509},
  eprint        = {2204.00200},
  primaryclass  = {hep-lat},
}

@article{LatticePartonLPC:2023pdv,
    author = "Chu, Min-Huan and others",
    collaboration = "Lattice Parton (LPC)",
    title = "{Lattice calculation of the intrinsic soft function and the Collins-Soper kernel}",
    eprint = "2306.06488",
    archivePrefix = "arXiv",
    primaryClass = "hep-lat",
    doi = "10.1007/JHEP08(2023)172",
    journal = "JHEP",
    volume = "08",
    pages = "172",
    year = "2023"
}

@article{Deng:2022gzi,
    author = "Deng, Zhi-Fu and Wang, Wei and Zeng, Jun",
    title = "{Transverse-momentum-dependent wave functions and soft functions at one-loop in large momentum effective theory}",
    eprint = "2207.07280",
    archivePrefix = "arXiv",
    primaryClass = "hep-th",
    doi = "10.1007/JHEP09(2022)046",
    journal = "JHEP",
    volume = "09",
    pages = "046",
    year = "2022"
}

@article{Bollweg:2025ecn,
    author = "Bollweg, Dennis and Gao, Xiang and Mukherjee, Swagato and Zhao, Yong",
    title = "{Lattice QCD Benchmark of Proton Helicity and Flavor-Dependent Unpolarized Transverse Momentum-Dependent Parton Distribution Functions at Physical Quark Masses}",
    eprint = "2505.18430",
    archivePrefix = "arXiv",
    primaryClass = "hep-lat",
    doi = "10.1103/tb2w-3tks",
    journal = "Phys. Rev. Lett.",
    volume = "135",
    number = "20",
    pages = "201901",
    year = "2025"
}

@article{Avkhadiev:2025wps,
    author = "Avkhadiev, Artur and Bertone, Valerio and Bissolotti, Chiara and Cerutti, Matteo and Fu, Yang and Rodini, Simone and Shanahan, Phiala and Wagman, Michael and Zhao, Yong",
    title = "{Extraction of the Collins-Soper Kernel from a Joint Analysis of Experimental and Lattice Data}",
    eprint = "2510.26489",
    archivePrefix = "arXiv",
    primaryClass = "hep-ph",
    reportNumber = "DESY-25-145, FERMILAB-PUB-25-0784-T, INT-PUB-25-026, MIT-CTP/5946",
    doi = "10.1103/zphz-4k8q",
    journal = "Phys. Rev. Lett.",
    volume = "136",
    number = "17",
    pages = "171902",
    year = "2026"
}

@article{Vermaseren:1982cz,
    author = "Vermaseren, J. A. M.",
    title = "{Two Photon Processes at Very High-Energies}",
    reportNumber = "NIKHEF-H/82-15",
    doi = "10.1016/0550-3213(83)90336-X",
    journal = "Nucl. Phys. B",
    volume = "229",
    pages = "347--371",
    year = "1983"
}

@inproceedings{Baranov:1991yq,
    author = "Baranov, S. P. and Duenger, O. and Shooshtari, H. and Vermaseren, J. A. M.",
    title = "{LPAIR: A generator for lepton pair production}",
    booktitle = "{Workshop on Physics at HERA}",
    year = "1991"
}

@article{Suri:1971yx,
    author = "Suri, Ashok and Yennie, Donald R.",
    title = "{The space-time phenomenology of photon absorption and inelastic electron scattering}",
    reportNumber = "SLAC-PUB-0954",
    doi = "10.1016/0003-4916(72)90242-4",
    journal = "Annals Phys.",
    volume = "72",
    pages = "243",
    year = "1972"
}

@article{Fernando:2025xzv,
    author = "Fernando, I. P. and Keller, D.",
    title = "{Deep neural network extraction of unpolarized transverse momentum distributions}",
    eprint = "2510.17243",
    archivePrefix = "arXiv",
    primaryClass = "hep-ph",
    doi = "10.1103/p3sg-k524",
    journal = "Phys. Rev. D",
    volume = "113",
    number = "9",
    pages = "096017",
    year = "2026"
}

@article{Hautmann:2019biw,
    author = "Hautmann, F. and Keersmaekers, L. and Lelek, A. and Van Kampen, A. M.",
    title = "{Dynamical resolution scale in transverse momentum distributions at the LHC}",
    eprint = "1908.08524",
    archivePrefix = "arXiv",
    primaryClass = "hep-ph",
    reportNumber = "CERN-TH-2019-130",
    doi = "10.1016/j.nuclphysb.2019.114795",
    journal = "Nucl. Phys. B",
    volume = "949",
    pages = "114795",
    year = "2019"
}

@article{Frederix:2018nkq,
    author = "Frederix, R. and Frixione, S. and Hirschi, V. and Pagani, D. and Shao, H. -S. and Zaro, M.",
    title = "{The automation of next-to-leading order electroweak calculations}",
    eprint = "1804.10017",
    archivePrefix = "arXiv",
    primaryClass = "hep-ph",
    reportNumber = "Nikhef/2018-015, TUM-HEP-1138/18, NIKHEF-2018-015, TUM-HEP-1138-18",
    doi = "10.1007/JHEP11(2021)085",
    journal = "JHEP",
    volume = "07",
    pages = "185",
    year = "2018",
    note = "[Erratum: JHEP 11, 085 (2021)]"
}

@article{Manohar:2016nzj,
    author = "Manohar, Aneesh and Nason, Paolo and Salam, Gavin P. and Zanderighi, Giulia",
    title = "{How bright is the proton? A precise determination of the photon parton distribution function}",
    eprint = "1607.04266",
    archivePrefix = "arXiv",
    primaryClass = "hep-ph",
    reportNumber = "CERN-TH-2016-155",
    doi = "10.1103/PhysRevLett.117.242002",
    journal = "Phys. Rev. Lett.",
    volume = "117",
    number = "24",
    pages = "242002",
    year = "2016"
}

@article{Kovarik:2015cma,
    author = "Kovarik, K. and others",
    title = "{nCTEQ15 - Global analysis of nuclear parton distributions with uncertainties in the CTEQ framework}",
    eprint = "1509.00792",
    archivePrefix = "arXiv",
    primaryClass = "hep-ph",
    reportNumber = "LPSC-15-153, MS-TP-15-11, FERMILAB-PUB-15-375-ND-PPD-T",
    doi = "10.1103/PhysRevD.93.085037",
    journal = "Phys. Rev. D",
    volume = "93",
    number = "8",
    pages = "085037",
    year = "2016"
}

@article{Hou:2016sho,
    author = "Hou, Tie-Jiun and others",
    title = "{Reconstruction of Monte Carlo replicas from Hessian parton distributions}",
    eprint = "1607.06066",
    archivePrefix = "arXiv",
    primaryClass = "hep-ph",
    reportNumber = "SMU-HEP-16-06",
    doi = "10.1007/JHEP03(2017)099",
    journal = "JHEP",
    volume = "03",
    pages = "099",
    year = "2017"
}

@article{H1:2015ubc,
    author = "Abramowicz, H. and others",
    collaboration = "H1, ZEUS",
    title = "{Combination of measurements of inclusive deep inelastic ${e^{\pm }p}$ scattering cross sections and QCD analysis of HERA data}",
    eprint = "1506.06042",
    archivePrefix = "arXiv",
    primaryClass = "hep-ex",
    reportNumber = "DESY-15-039",
    doi = "10.1140/epjc/s10052-015-3710-4",
    journal = "Eur. Phys. J. C",
    volume = "75",
    number = "12",
    pages = "580",
    year = "2015"
}

@article{Tan:2025ofx,
    author = "Tan, Jin-Xin and others",
    title = "{Lattice QCD determination of the Collins-Soper kernel in the continuum and physical mass limits}",
    eprint = "2511.22547",
    archivePrefix = "arXiv",
    primaryClass = "hep-lat",
    doi = "10.1103/pry5-7729",
    journal = "Phys. Rev. D",
    volume = "113",
    number = "5",
    pages = "054505",
    year = "2026"
}

@article{Francis:2026czb,
    author = "Francis, Anthony and Lin, C. -J. David and Morris, Wayne and Zhao, Yong",
    title = "{The Collins-Soper kernel from a vacuum soft function}",
    eprint = "2606.19221",
    archivePrefix = "arXiv",
    primaryClass = "hep-lat",
    month = "6",
    year = "2026"
}

@article{Avkhadiev:2026xyf,
    author = "Avkhadiev, Artur and Fu, Yang and Shanahan, Phiala E. and Wagman, Michael L. and Zhao, Yong",
    title = "{First constraints on the nonperturbative gluon Collins-Soper kernel}",
    eprint = "2607.24587",
    archivePrefix = "arXiv",
    primaryClass = "hep-lat",
    reportNumber = "FERMILAB-PUB-26-0391-T, MIT-CTP/6053, INT-PUB-26-024",
    month = "7",
    year = "2026"
}

@article{Deak:2011ga,
    author = "Deak, M. and others",
    title = "{Forward Jets and Energy Flow in Hadronic Collisions}",
    eprint = "1112.6354",
    archivePrefix = "arXiv",
    primaryClass = "hep-ph",
    reportNumber = "DESY-11-262, IFJPAN-IV-2011-8, OUTP-11-38-P",
    doi = "10.1140/epjc/s10052-012-1982-5",
    journal = "Eur. Phys. J. C",
    volume = "72",
    pages = "1982",
    year = "2012"
}

@article{Deak:2010gk,
    author = "Deak, M. and others",
    title = "{Forward-Central Jet Correlations at the Large Hadron Collider}",
    eprint = "1012.6037",
    archivePrefix = "arXiv",
    primaryClass = "hep-ph",
    reportNumber = "DESY-10-179, OUTP-10-07-P",
    month = "12",
    year = "2010"
}

@article{Deak:2009xt,
    author = "Deak, M. and others",
    title = "{Forward Jet Production at the Large Hadron Collider}",
    eprint = "0908.0538",
    archivePrefix = "arXiv",
    primaryClass = "hep-ph",
    reportNumber = "DESY-09-119, OUTP-09-15P",
    doi = "10.1088/1126-6708/2009/09/121",
    journal = "JHEP",
    volume = "09",
    pages = "121",
    year = "2009"
}

@article{Hautmann:2007yok,
    author = "Hautmann, F. and Jung, H.",
    editor = "Catani, Stefano and Colferai, Dimitri and Grazzini, Massimiliano",
    title = "{Three-jet DIS final states from k(T)-dependent parton showers}",
    eprint = "0804.1746",
    archivePrefix = "arXiv",
    primaryClass = "hep-ph",
    reportNumber = "OUTP-08-04-P",
    doi = "10.22323/1.048.0030",
    journal = "PoS",
    volume = "RADCOR2007",
    pages = "030",
    year = "2007"
}

@article{ATLAS:2015iiu,
    author = "Aad, Georges and others",
    collaboration = "ATLAS",
    title = "{Measurement of the transverse momentum and $\phi ^*_{\eta }$ distributions of Drell{\textendash}Yan lepton pairs in proton{\textendash}proton collisions at $\sqrt{s}=8$  TeV with the ATLAS detector}",
    eprint = "1512.02192",
    archivePrefix = "arXiv",
    primaryClass = "hep-ex",
    reportNumber = "CERN-PH-EP-2015-275",
    doi = "10.1140/epjc/s10052-016-4070-4",
    journal = "Eur. Phys. J. C",
    volume = "76",
    number = "5",
    pages = "291",
    year = "2016"
}

@article{Cole:2026eex,
    author = "Cole, Ella and Costantini, Mark N. and Hammou, Elie and Mantani, Luca and Merlotti, Francesco and Morales-Alvarado, Manuel and Ubiali, Maria",
    title = "{Tailored PDFs for new physics searches}",
    eprint = "2602.20235",
    archivePrefix = "arXiv",
    primaryClass = "hep-ph",
    doi = "10.1007/JHEP07(2026)043",
    journal = "JHEP",
    volume = "07",
    pages = "043",
    year = "2026"
}

@article{Fiaschi:2022wgl,
    author = "Fiaschi, J. and Giuli, F. and Hautmann, F. and Moch, S. and Moretti, S.",
    title = "{Z'-boson dilepton searches and the high-x quark density}",
    eprint = "2211.06188",
    archivePrefix = "arXiv",
    primaryClass = "hep-ph",
    reportNumber = "CERN-TH-2022-177, LTH 1323, DESY-22-178",
    doi = "10.1016/j.physletb.2023.137915",
    journal = "Phys. Lett. B",
    volume = "841",
    pages = "137915",
    year = "2023"
}

@article{Fu:2023rrs,
    author = "Fu, Yao and Brock, Raymond and Hayden, Daniel and Yuan, Chien-Peng",
    title = "{Probing parton distribution functions at large x via Drell-Yan forward-backward asymmetry}",
    eprint = "2307.07839",
    archivePrefix = "arXiv",
    primaryClass = "hep-ph",
    doi = "10.1103/PhysRevD.109.054006",
    journal = "Phys. Rev. D",
    volume = "109",
    number = "5",
    pages = "054006",
    year = "2024"
}

@article{Bertone:2024snr,
    author = "Bertone, Valerio and Bozzi, Giuseppe and Hautmann, Francesco",
    title = "{Perturbative RGE systematics in precision observables}",
    eprint = "2407.20842",
    archivePrefix = "arXiv",
    primaryClass = "hep-ph",
    doi = "10.1103/PhysRevD.111.074005",
    journal = "Phys. Rev. D",
    volume = "111",
    number = "7",
    pages = "074005",
    year = "2025"
}

@article{Taels:2022tza,
    author = "Taels, Pieter and Altinoluk, Tolga and Beuf, Guillaume and Marquet, Cyrille",
    title = "{Dijet photoproduction at low x at next-to-leading order and its back-to-back limit}",
    eprint = "2204.11650",
    archivePrefix = "arXiv",
    primaryClass = "hep-ph",
    doi = "10.1007/JHEP10(2022)184",
    journal = "JHEP",
    volume = "10",
    pages = "184",
    year = "2022"
}

@article{Caucal:2023fsf,
    author = {Caucal, Paul and Salazar, Farid and Schenke, Bj{\"o}rn and Stebel, Tomasz and Venugopalan, Raju},
    title = "{Back-to-Back Inclusive Dijets in Deep Inelastic Scattering at Small x: Complete NLO Results and Predictions}",
    eprint = "2308.00022",
    archivePrefix = "arXiv",
    primaryClass = "hep-ph",
    doi = "10.1103/PhysRevLett.132.081902",
    journal = "Phys. Rev. Lett.",
    volume = "132",
    number = "8",
    pages = "081902",
    year = "2024"
}

@article{Altinoluk:2023hfz,
    author = "Altinoluk, Tolga and Armesto, Nestor and Kovner, Alexander and Lublinsky, Michael",
    title = "{Single inclusive particle production at next-to-leading order in proton-nucleus collisions at forward rapidities: Hybrid approach meets TMD factorization}",
    eprint = "2307.14922",
    archivePrefix = "arXiv",
    primaryClass = "hep-ph",
    doi = "10.1103/PhysRevD.108.074003",
    journal = "Phys. Rev. D",
    volume = "108",
    number = "7",
    pages = "074003",
    year = "2023"
}

@article{Wang:2022zdu,
    author = "Wang, Lei and Chen, Lin and Gao, Zhan and Shi, Yu and Wei, Shu-Yi and Xiao, Bo-Wen",
    title = "{Forward inclusive jet productions in pA collisions}",
    eprint = "2211.08322",
    archivePrefix = "arXiv",
    primaryClass = "hep-ph",
    doi = "10.1103/PhysRevD.107.016016",
    journal = "Phys. Rev. D",
    volume = "107",
    number = "1",
    pages = "016016",
    year = "2023"
}

@article{Deganutti:2023qct,
    author = "Deganutti, Federico and Royon, Christophe and Schlichting, Soeren",
    title = "{Forward dijet production at the LHC within an impact parameter dependent TMD approach}",
    eprint = "2311.01965",
    archivePrefix = "arXiv",
    primaryClass = "hep-ph",
    doi = "10.1007/JHEP01(2024)159",
    journal = "JHEP",
    volume = "01",
    pages = "159",
    year = "2024"
}

@article{vanHameren:2023oiq,
    author = "van Hameren, A. and Kakkad, H. and Kotko, P. and Kutak, K. and Sapeta, S.",
    title = "{Searching for saturation in forward dijet production at the LHC}",
    eprint = "2306.17513",
    archivePrefix = "arXiv",
    primaryClass = "hep-ph",
    doi = "10.1140/epjc/s10052-023-12120-7",
    journal = "Eur. Phys. J. C",
    volume = "83",
    number = "10",
    pages = "947",
    year = "2023"
}

@article{Caucal:2025zkl,
    author = {Caucal, Paul and Kang, Zhong-Bo and Korcyl, Piotr and Salazar, Farid and Schenke, Bj{\"o}rn and Stebel, Tomasz and Venugopalan, Raju and Zhao, Wenbin},
    title = "{Probing gluon saturation with forward di-hadron correlations in proton-nucleus collisions}",
    eprint = "2512.21466",
    archivePrefix = "arXiv",
    primaryClass = "hep-ph",
    doi = "10.1016/j.physletb.2026.140599",
    journal = "Phys. Lett. B",
    volume = "879",
    pages = "140599",
    year = "2026"
}

@article{Motyka:2016lta,
    author = "Motyka, Leszek and Sadzikowski, Mariusz and Stebel, Tomasz",
    title = "{Lam-Tung relation breaking in $Z^0$ hadroproduction as a probe of parton transverse momentum}",
    eprint = "1609.04300",
    archivePrefix = "arXiv",
    primaryClass = "hep-ph",
    doi = "10.1103/PhysRevD.95.114025",
    journal = "Phys. Rev. D",
    volume = "95",
    number = "11",
    pages = "114025",
    year = "2017"
}

@article{Bacchetta:2022awv,
    author = "Bacchetta, Alessandro and Bertone, Valerio and Bissolotti, Chiara and Bozzi, Giuseppe and Cerutti, Matteo and Piacenza, Fulvio and Radici, Marco and Signori, Andrea",
    collaboration = "MAP (Multi-dimensional Analyses of Partonic distributions)",
    title = "{Unpolarized transverse momentum distributions from a global fit of Drell-Yan and semi-inclusive deep-inelastic scattering data}",
    eprint = "2206.07598",
    archivePrefix = "arXiv",
    primaryClass = "hep-ph",
    doi = "10.1007/JHEP10(2022)127",
    journal = "JHEP",
    volume = "10",
    pages = "127",
    year = "2022"
}

@article{Xiao:2026tbs,
    author = "Xiao, Bo-Wen and Zhao, Yuxiang and Zhou, Jian",
    title = "{Physics of the Electron-Ion Collider in China}",
    eprint = "2608.11712",
    archivePrefix = "arXiv",
    primaryClass = "hep-ph",
    doi = "10.1016/j.ppnp.2026.104264",
    journal = "Prog. Part. Nucl. Phys.",
    volume = "151",
    pages = "104264",
    year = "2026"
}

@article{Li:2026uug,
    author = "Li, Wanchen and Shao, Ding Yu and Wei, Shu-Yi and Zhou, Jian",
    title = "{Recoil Geometry Unmasks Gluon Saturation in Forward $Z^0$ Production}",
    eprint = "2609.02356",
    archivePrefix = "arXiv",
    primaryClass = "hep-ph",
    month = "9",
    year = "2026"
}

\end{document}